\documentclass{aa}  
\usepackage{graphicx}
\usepackage{natbib}
\usepackage[figuresright]{rotating}
\usepackage{txfonts}
\def\magarc{mag\,arcsec$^{-2}$}
\def\2{$^{2}$}
\def\1{$^{-1}$}
\def\3{$^{-3}$}
\def\rad{$_{rad}$}
\def\m{$\mu$}

\def\si{$\sim$}

\def\loweq{$\le$}
\def\lareq{$\ge$}

\def\sm{$\sim$}

\def\II{{\sc ii}}
\def\III{{\sc iii}}
\def\B{$\beta$}
\def\A{$\alpha$}
\def\La{$\lambda$}

\def\plmi{$\pm$}
\def\sun{$_{\odot}$}

\def\Ph{$\Phi$}
\def\sig{$\sigma$}

\begin{document}
   \title{Low Surface Brightness Galaxies around the HDF-S}

   \subtitle{II. Distances and volume densities}

   \author{L. Haberzettl
          \inst{1,2}
          \and
          D.J. Bomans
          \inst{1}
          \and
          R.-J. Dettmar
          \inst{1}
          }

   \offprints{L. Haberzettl}

   \institute{Astronomical Institute Ruhr-University Bochum,
              Universit\"atsstrasse 150, 44780 Bochum, Germany
              \and
              present address: Department of Physics and Astronomy, University
              of Louisville, Louisville, KY, 40292, USA\\
              \email{lghabe01@louisville.edu}
}
   \date{}

 
  \abstract
   {}
   {With this study we aim at the spectroscopic verification of a
     photometrically selected 
     sample of Low Surface Brightness (LSB) galaxy candidates in a field
     around the Hubble Deep Field-South (HDF-S). The sample helps to extend the
     parameter space for LSB galaxies to lower central surface
     brightnesses 
     and to provide better estimates on the volume densities of these objects. 
}
   {To derive redshifts for the LSB candidates, long-slit spectra were
    obtained covering a 
    spectral range from 3400\,\AA\, to 7500 \AA. The observations have been
    obtained using the ESO 3.6\,m telescope, equipped with the EFOSC2
    spectrograph. From the measured radial velocities, distances could be
    estimated. With this 
    distance information, it is possible to differentiate between true LSB
    galaxies and higher redshift High Surface Brightness (HSB) galaxies
    which may contaminate the
    sample. A correction for the surface brightnesses 
    can then be applied, accounting for the cosmological dimming effect
    (``Tolman Dimming''). 
}
   {We show that \si70\% of the LSB candidates, selected based on their
    location in the color-color space, are real LSB galaxies.  
   Their position in the color-color diagrams, therefore, indicate 
   that the LSB galaxies have a different stellar population mix resulting
   from a different star formation history compared to HSBs. Our LSB
    galaxy sample consists only of large disk galaxies with scale-length
    between 2.5\,kpc and 7.3\,kpc. We confirm the flat central surface
    brightness distribution of  previous surveys and extend this distribution
    down to central surface brightnesses of 27\,$B$\,\magarc. 
}
   {}

   \keywords{Surveys -- galaxies: distances and redshifts -- galaxies:
   fundamental parameters (luminosities, radii)}

   \maketitle
%

\section{Introduction}
The known properties of LSB galaxies are still a challenge for existing
  theories of galaxy formation and evolution. 
Although they may not represent a significant amount of the
luminosity density in the universe, their number density is very
high. Searches for LSB galaxies showed that they account for up to 60\,\% of
the local galaxy population \citep{1996MNRAS.280..337M,2000ApJ...529..811O}.
\citet{2004MNRAS.355.1303M} showed in their study of the bivariate
surface brightness distribution, that up to 20\,\% of the dynamical
mass is represented by the LSB galaxies and that they account for up to 60\,\%
of the number density in the Universe.
Thus it seems, that LSB galaxies are a common product of galaxy
formation and evolutionary processes. Therefore, it is essential to understand
their role in formation and evolution scenarios of galaxies in general.

Studies of the properties of local LSB galaxy samples have
shown that they populate nearly the whole parameter space derived also for
HSB galaxies. The only difference is their low
central surface brightness which is below $\mu_0$\,=\,22.5\,mag\,arcsec$^{-2}$
in the $B$-filter.
Field LSB galaxies are generally gas rich, although the gas surface densities
are very low, too. The typical gas surface densities for LSB galaxies are below
the Kennicutt criterion for ongoing star formation
\citep{1989ApJ...344..685K,1993AJ....106..548V,1997AJ....114.1858P} which
results in a suppressed current star formation rate. The morphology of LSB
galaxies does not show any significant differences to the morphology of HSB
galaxies. The properties of LSBs are independent of the Hubble type, these
galaxies exist over the whole Hubble sequence \citep{1992AJ....103.1107S}. 
While the LSB galaxies could be found in all morphological types and
  results are comparable to results from surveys governed by HSB galaxies,
  e.g. UGC catalog \citep{1973ugcg.book.....N}, the major part of the
  population consists of late-type galaxies. Elliptical LSB
  galaxies are rare in these surveys (\sm\,14\%) and found mostly in cluster
  environment, which means they underwent a different star formation history
  than field LSB galaxies.

The simplest evolutionary scenario suggests that LSB galaxies are faded
remnants of HSB galaxies. For this scenario one would expect that LSB galaxies
are found to have red colors. This scenario could be ruled out by the fact
that LSB galaxies are found to cover the whole color space from
very red to very blue. Most of the LSB galaxies have quite blue colors which
could be explained only partly as the result of the relatively low
metalicities
\citep[Z\,$<$\,1/3\,Z$_{\odot}$,e.g.][]{1995A&A...302..353R,1993AAS...183.5704M}.
\citet{1995MNRAS.274..235D_B} showed that their sample of 20 LSB galaxies tend
to have bluer colors compared to the mean values estimated for HSB galaxies
of the corresponding Hubble type. This effect is mainly seen in the $B\,-\,R$
color (HSB: $\overline{B\,-\,R}\,=\,0.92$, LSB: $\overline{B\,-\,R}\,=\,0.78$)
and the $V\,-\,I$ color (HSB: $\overline{V\,-\,I}\,=\,0.90$, LSB:
$\overline{V\,-\,I}\,=\,0.76$).
These blue colors are a first hint for relatively young luminosity
  weighted average ages 
of the stellar populations in the LSB galaxies. However, until now we do have
only little information about the evolutionary paths taken by the LSB
galaxies, resulting in such low surface brightnesses 
  \citep[e.g.][]{1999A&A...342..655G,2000MNRAS.312..470B,2000A&A...357..397V}.

In this paper we present spectroscopic results for the LSB candidate sample
derived in a field around the HDF-S \citep{2007A&A...465...95H}.  We
used spectroscopic measurements of redshifted emission lines to derive
distances for these objects.
From the distances we calculate physical parameters of the derived galaxies and
distinguish the real LSB galaxies from the higher redshifted background
objects. We also discuss the behavior of the LSB galaxies in the color-color
space. Finally we compare our findings to results of previous searches
for LSB galaxies.

\section{Observation and data reduction}
\label{specfollow}
\subsection{Observations}
\label{hdfs_lsbs}
  We performed spectroscopic observations of a sample of 9 LSB galaxy
  candidates during the nights of October 23rd-25th 2000 at the ESO 3.6\,m
  telescope on La Silla. The spectra were obtained using the ESO Faint Object
  Spectrograph 2 (EFOSC2) equipped with a Loral 2048\2 pixel\2 CCD
  and a pixel
  size of 15\,\m m (0.157\,arcsec). 
  We took long-slit spectra using a grism of 300 gr\,mm\1 (grism \# 11),
  resulting in a spectral sampling of 2.12\,\AA\,pixel\1. With this setup we
  were able to cover a 
  spectral range from 3400\,\AA\, to 7500\,\AA. The CCD was binned by a factor
  of 2 in both directions. This setup resulted in a spectral resolution of
  $\sim$\,4\,\AA\,pixel\1. The spectrograph was equipped with a long-slit of
  slightly less than 5 arcmin length in the spatial direction. For the
  observation we chose a slit width of 1.2\,arcsec. This led to a FWHM for
  emission lines of \si\,15.6\,\AA\, at 4000\,\AA\, in the derived spectra.

  \quad
  In order to minimize flux losses due to atmospheric differential refraction
  \citep{1982PASP...94..715F}, the observations were carried out in an airmass
  range between 1.1 and 1.8. This resulted in a refraction below 1.8\,arcsec
  at 3500\,\AA\, and above $-$0.9\,arcsec at 7500\,\AA. Due to the sizes of the
  observed objects (\si\,10 to \si\,30 arcsec) these are still acceptable
  values. During the nights of the 23rd and 24th, the observing conditions
  were not photometric with a mean
  seeing of \si\,2\,arcsec. In the last night, the conditions were nearly
  photometric with a seeing below 1\,arcsec. Bias frames, dome flats,
  twilight-sky flats, and He-Ne comparison lamp exposures were taken at the
  beginning and the end of each night. For flux calibration, spectroscopic
  standard stars \citep{1992PASP..104..533H} were observed several times
  during each night. For the observations of the standard stars, we
  used the maximum slit width of 5\,arcsec in order to increase the signal to
  noise ratio of the spectra as much as possible. Each science exposure was
  obtained with an exposure time of about 1800\,s. In most cases, two
  exposures per object were observed (see Table~\ref{object}). 
\begin{table}[h]
\centering
\begin{tabular}{c c c}
\hline
\hline
NAME&EXPOSURES TIME [s]&EXPOSURES\\
\hline
\object{LSB J22311-60503}& 1800& 2\\
\object{LSB J22324-60520}& 1800& 2\\
\object{LSB J22325-60155}& 1800& 2\\
\object{LSB J22330-60543}& 1800& 2\\
\object{LSB J22343-60222}& 1800& 2\\
\object{LSB J22352-60420}& 1800& 2\\
\object{LSB J22353-60311}& 1800& 1\\
\object{LSB J22353-60122}& 1800& 2\\
\object{LSB J22355-60183}& 1800& 2\\
\hline
\end{tabular}
\caption{Exposure times and number of exposures for
  the HDF-S LSB candidate sample observed with ESO 3.6\,m telescope.}
\label{object}
\end{table}

\subsection{Reduction and calibration}
\label{redu_cal}
  The data reduction and calibration were carried out based on the standard
  reduction procedures within IRAF \citep{massey97, massey92}. 
  The wavelength calibration resulted in a spectral sampling of 4.04\,\AA\,
  per pixel with an rms for the wavelength fit between 0.3 
  \AA\, and 1.0 \AA\, per pixel. After the wavelength calibration, we combined
  the single science exposures of every object. This coaddition makes it also
  possible to remove cosmic rays from the science spectra. The cosmic ray
  cleaning of LSB J22353-60311, for which we only observed one spectra,
  was done using the cosmic ray identification task L.A. cosmic
  \citep{2001PASP..113.1420V}. This cosmic ray cleaning is based on a Laplace
  filtering method. In a next step we flux calibrated the coadded and cosmic
  ray cleaned spectra using the observed standard star spectra. Finally, we
  applied an extinction correction to the science spectra. 
  We used only the foreground
  extinction value of $E(B-V)\,=\,0.028$ for this correction, which we adopted
  from the extinction maps of \citet{1998ApJ...500..525S}.  Applying only
  foreground extinction from the galaxy gives reasonable results for the LSB
  galaxies, since internal extinction can be neglected for these galaxies due
  to the low dust content. 
  There exists only a few week detections for LSB galaxies in the FIR
  and mm-wavelength region where dust radiates thermal emission
  \citep{1994AJ....108..446H}. The final result of the reduction and
  calibration processes are 2D- and 1D-spectra which will be used in the
  following analysis.
\begin{figure*}[ht]
\centering
\includegraphics[width=7.5cm]{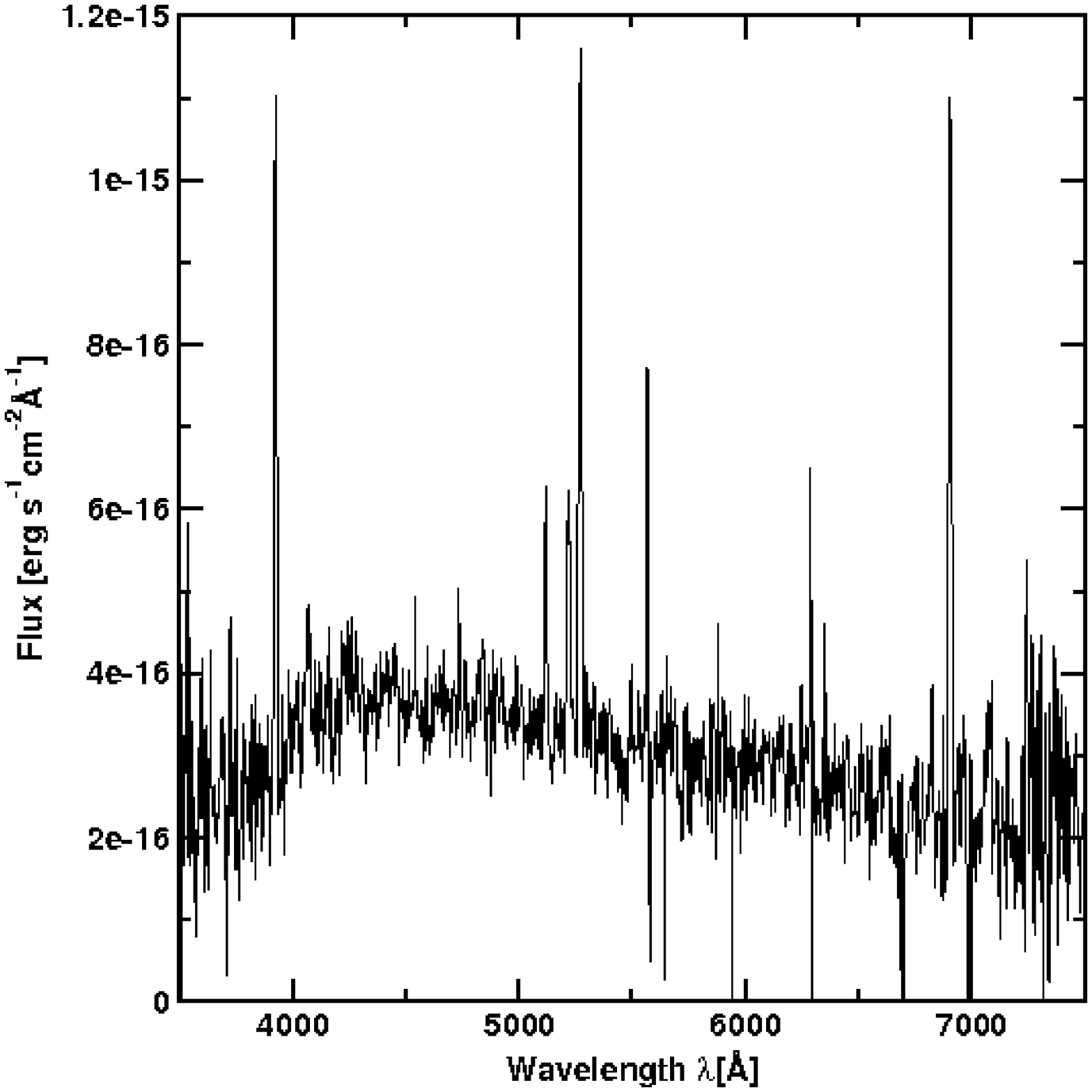}
\includegraphics[width=8cm]{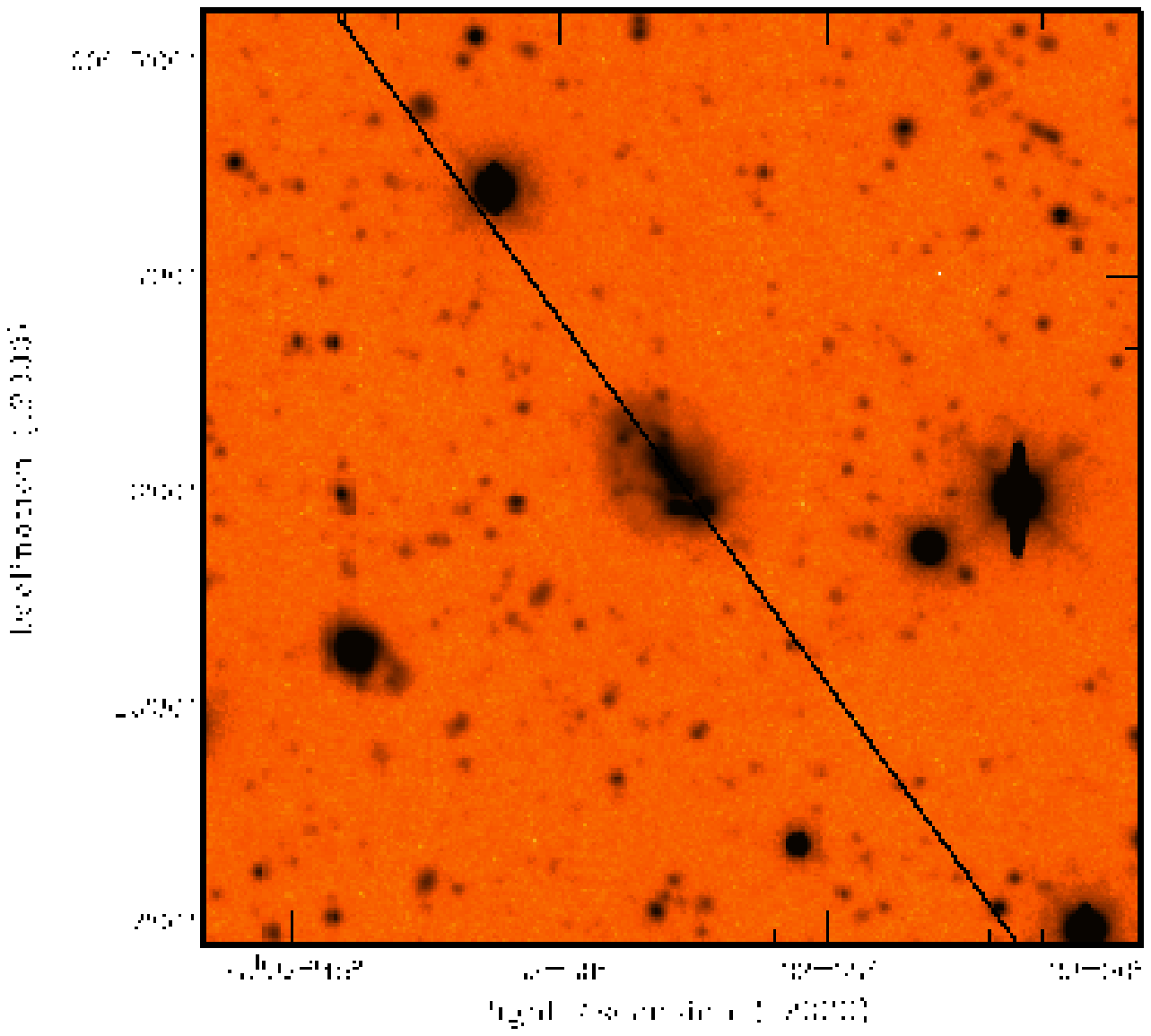}
\caption{In the following we show the 1D spectra and images of the
  spectroscopically observed LSB galaxy candidates in the HDF-S
  (e.g. \object{LSB J22325-60155}). The
  orientation and location of the slit is stated by the black line in the
  $2^\prime\times2^\prime$ large images.}
\label{spectra}
\end{figure*}

\section{Results and Discussion}
\subsection{Distances and Physical Parameters of the LSB Sample} 
\label{secdist}
\begin{table*}
\centering
\begin{tabular}{l l l l l l l l c c}
\hline
\hline 
NAME&[O\II]&H\B&[O\III]&[O\III]&H\A&[N\II]&[S\II]&z&$\sigma_z$\\
&\La3727&\La4861&\La4959&\La5007&\La6563&\La6584&\La\La6717,6731&&\\
&[\AA]&[\AA]&[\AA]&[\AA]&[\AA]&[\AA]&[\AA]&&\\
\hline
\object{LSB J22311-60503}&4172&--&--&-&7351&--&--&0.1198&0.0004\\
\object{LSB J22325-60155}&3924&5118&5219&5270&6907&--&7077$^{a}$&0.0526&0.0002\\
\object{LSB J22330-60543}&3936&5133&5236&5287&6929&--&7096$^{a}$&0.0559&0.0001\\
\object{LSB J22343-60222}&3903&5089&5190&5242&6866&6888$^{a}$&7037$^{a}$&0.0467&0.0005\\
\object{LSB J22352-60420}&4170&5439&5550$^{a}$&5603$^{a}$&7347&--&--&0.1191&0.0003\\
\object{LSB J22353-60311}&3946&5146&5250&5297&6943&6968$^{a}$&--&0.0584&0.0004\\
\object{LSB J22354-60122}&4329&5647&5758&5816&--&--&--&0.1615&0.0002\\
\object{LSB J22355-60183}&4194&5468&5584$^{a}$&5632&7381&--&--&0.1249&0.0003\\
\hline
\object{LSB J22324-60520}&--&--&--&--&--&--&--&0.1723$^{b}$&--\\
\hline
\end{tabular}
\caption{Measurements for the emission lines of the
  LSB candidate sample. All entries which are marked with $^{a}$ were not used
  for the estimation of the mean redshift $<z>$ and therefore also not for
  the estimation of the standard deviation $\sigma_z$. All lines which were
  either influenced by night sky emission lines or which had only upper limits
  were also not used for the determination of the mean redshift. It was not
  possible to resolve the two component of the [S\II]-doublet. Therefore
  [S\II] was not used in order to estimate redshifts. The redshift of
  \object{LSB J22324-60520} (marked with $^{b}$) was estimated measuring the
  wavelength of the Ca\II-K absorption line.}
\label{lines}
\end{table*}

  Most of the common internal parameters of the studied galaxies e.g., sizes
  in physical units (radii, scale-length), luminosities, absolute magnitudes
  (see Table~\ref{physpar}) are distance-dependent parameters. The radii are
  measured by eye, following the light distribution into the noise. This means
  that the radius is measured at the limiting surface brightness of
  \si\,29\,\magarc\, \citep[see][]{2007A&A...465...95H}. We chose the
  radii at the 
  limiting surface brightness, because the Holmberg radius at 25\,\magarc\,
  would give no useful information about the sizes of LSB galaxies due
  to the low surface brightnesses in these galaxies. In order to
  derive the distances, we estimated the redshifts z for our sample galaxies
  as the mean of the redshifts calculated for every well detected emission
  line in the single galaxy spectra. 
  
  \quad
  In order to maximize the signal-to-noise ratios we summed up all columns
  along the spatial axis which contributed a significant signal to the final
  science spectra. In these integrated 1D spectra (see Fig.~\ref{spectra}), it
  was now possible to detect the most important emission lines e.g., [O\II],
  H\,\B, [O\III]-doublet, and H\,\A. In some cases, we also detected the
  [S\,\II]-doublet and derived useful upper limits for the [N\,\II]\La6584\,
  emission line.

  \quad
  For the determination of the mean redshift, we used only emission lines which
  were clearly detected and which were not influenced by night sky emission
  lines. The lines which are only measured as upper limits or which were
  influenced by night sky emission lines are marked with $^{a}$ in
  Table~\ref{lines}. Due to the low resolution, it was not possible to resolve
  the [S\II]\La\La6717,6731-doublet. Therefore, we also did not use this
  doublet for the estimation of the mean redshift and marked them with
  $^{a}$ in Table~\ref{lines}. The errors of the mean redshifts were
  calculated using the standard deviation.

  \quad
  Following the formalism in Eqn.~\ref{hubbellaw}-\ref{luminos}, we calculated
  physical parameters (see Table~\ref{physpar}) for the observed galaxy
  sample, using the estimated redshifts e.g., radial velocities v\rad, proper
  distances D, radii r in physical units (proper length), scale-length h in
  physical units (proper length), absolute $B$-band magnitudes M$_B$ and
  luminosities L$_B$.  
  Eqn.~\ref{hubbellawred} and \ref{prodist} were used for galaxies having
  distances z\,\lareq\,0.1:  
\begin{eqnarray}
\label{hubbellaw}
\mathrm{v_{rad}}[z<0.1]&=&c\cdot z \\
\label{hubbellawred}
\mathrm{v_{rad}}[z\ge0.1]&=&c\cdot \frac{(1+z)^2-1}{(1+z)^2+1}\\
\label{propperdist}
D[z<0.1]&=&\frac{\mathrm{v_{rad}}}{H_0}\\
D[z\ge 0.1]&=&\frac{2c}{H_0}\frac{1}{1+z}\left[\left(1+z\right)-\left(1+z\right)^{1/2} \right]
\label{prodist}
\end{eqnarray} 
  For the calculations we assumed a Hubble constant $H_0\,=\,71$ kms\1Mpc\3
  \citep{2003ApJS..148..175S}: 
\begin{eqnarray}
r[z<0.1]&=&2\cdot D \cdot \tan\frac{r_\mathrm{arcsec}}{2}\\
h[z<0.1]&=&2\cdot D\cdot \tan\frac{\alpha_{B_\mathrm{W}}}{2}\\
\label{distcos1}
r[z\ge0.1]&=&d=\frac{r\mathrm{[rad]}\cdot D}{1+z}\\
h[z\ge 0.1]&=&d=\frac{\alpha_{B_\mathrm{W}}\mathrm{[rad]}\cdot D}{1+z}
\label{distcos2}
\end{eqnarray}
  The radius r (proper length d) in physical units as well as the scale-length
  h (proper length) in physical units of galaxies with redshifts above
  $z\,>\,0.1$ were calculated using Eqns.~\ref{distcos1} \& \ref{distcos2}
  \citep{longair}. The absolute magnitudes result from the distance modulus 
  \citep[see Eqn.~\ref{distancemodule};][]{1994MNRAS.266..155D} using the
  proper distance D (see Eqn.~\ref{propperdist} \& \ref{prodist}). We used the
  absolute magnitudes to estimate the $B$-band luminosities $L_B$ in units of
  solar luminosities (see Eqns.~\ref{luminos}): 
\begin{eqnarray}
\label{distancemodule}
M_{\nu}=m_{\nu}-5\cdot\log_{10}\left[\frac{2\cdot c}{H_0}\cdot
  \left((1+z)-(1+z)^{1/2}\right)\right]\\
\nonumber
-25-2.5\cdot z \\
M_{B}=4.87-2.5\cdot\log\left(\frac{L}{L_{\odot}}\right)
\label{luminos}
\end{eqnarray}
  We derived the physical parameters for the whole HDF-S galaxy sample,
  including \object{LSB J22324-60520} for which we detected no emission
  lines. This 
  galaxy shows a spectrum of an elliptical galaxy with no ongoing star
  formation. Therefore, we estimated the redshift by measuring the wavelength
  of the CaII-K absorption line and comparing this value to the rest frame 
  wavelength (\La$_0\,=\,3933\,\AA$). In the next sections we will discus the
  properties of the spectroscopically verified LSB galaxies found in the
  HDF-S.
\setlength{\tabcolsep}{1.5mm}
\begin{table*}
\centering
\begin{tabular}{l l l l l l l l l l}
\hline
\hline 
NAME&RA&DEC&v\rad&d&r$_m$&r&h&$M_B$&$L_B$\\
&[hh:mm:ss.s]&[dd:mm:ss]&[km s\1]&[Mpc]&[arcsec]&[kpc]&[kpc]&[mag]&[10$^9\,\cdot\,$L\sun]\\
\hline
\object{LSB J22311-60503}&22:31:13.0&-60:50:34&33854\plmi78&503.7\plmi25.1&12.9&31.5\plmi1.6&6.2\plmi0.3&-20.36\plmi0.07&12.36\\
\object{LSB J22324-60520}&22:32:41.8&-60:52:07&46717&638.0&8.5&7.15&1.25&-20.17&10.38\\
\object{LSB J22325-60155}&22:32:52.2&-60:15:58&15780\plmi69&221.2\plmi11.0&17.2&18.4\plmi0.9&4.6\plmi0.2&-16.90\plmi0.09& 0.51\\
\object{LSB J22330-60543}&22:33:03.2&-60:54:38&16770\plmi33&235.0\plmi11.7&15.1&17.1\plmi0.9&2.5\plmi0.1&-18.31\plmi0.08& 1.87\\
\object{LSB J22343-60222}&22:34:32.3&-60:22:21&14010\plmi159&196.4\plmi10.0&13.8&13.1\plmi0.7&4.3\plmi0.1&-17.43\plmi0.09& 0.83\\
\object{LSB J22352-60420}&22:35:22.5&-60:42:09&33615\plmi62&500.8\plmi24.9&10.8&23.4\plmi1.2&6.6\plmi0.3&-18.67\plmi0.15& 2.61\\
\object{LSB J22353-60311}&22:35:34.4&-60:31:12&17520\plmi120&245.6\plmi12.3&10.8&12.9\plmi0.6&7.3\plmi0.4&-17.77\plmi0.11& 1.14\\
\object{LSB J22354-60122}&22:35:46.1&-60:12:20&44581\plmi44&679.0\plmi33.8&7.6&21.5\plmi1.1&6.2\plmi0.3&-19.82\plmi0.13& 7.52\\
\object{LSB J22355-60183}&22:35:58.3&-60:18:40&35149\plmi57&525.2\plmi26.1&7.8&17.7\plmi0.9&5.3\plmi0.3&-19.15\plmi0.13& 4.06\\
\hline
\end{tabular}
\caption{Measured (e.g., r$_m$ measured radii) and calculated
  parameters of the observed HDF-S galaxy sample. The parameters in
  columns 4, 5, 7, 8, 9 and 10 are calculated using Eqns.~\ref{hubbellaw} to
  \ref{luminos} (see also description in Section~\ref{secdist}).}
\label{physpar}
\end{table*}

\subsection{The HDF-S LSB Galaxies in the Color-Color Space}
\label{color_space}
\begin{figure*}[ht]
\begin{center}
\includegraphics[width=8cm]{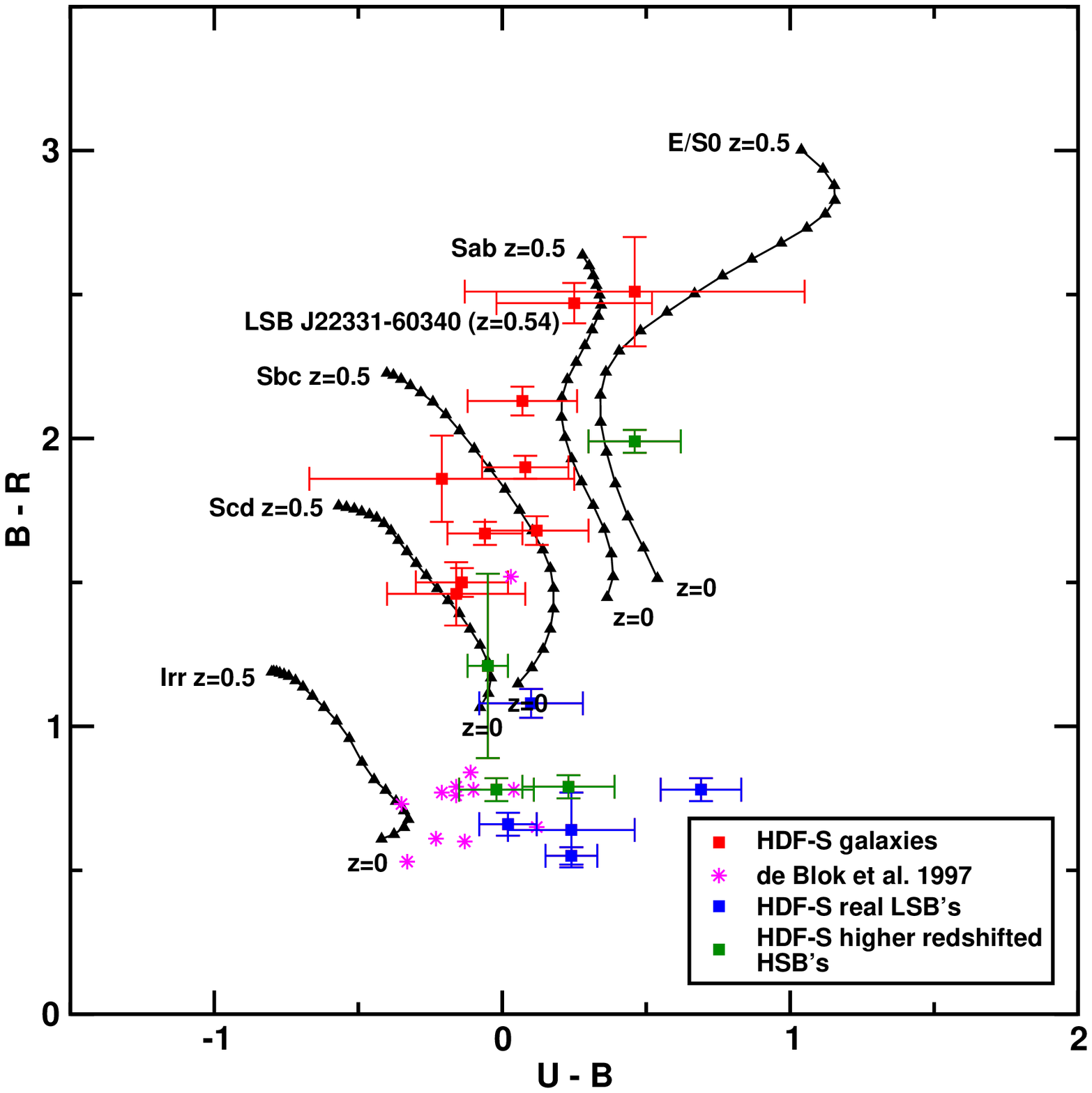}
\includegraphics[width=8cm]{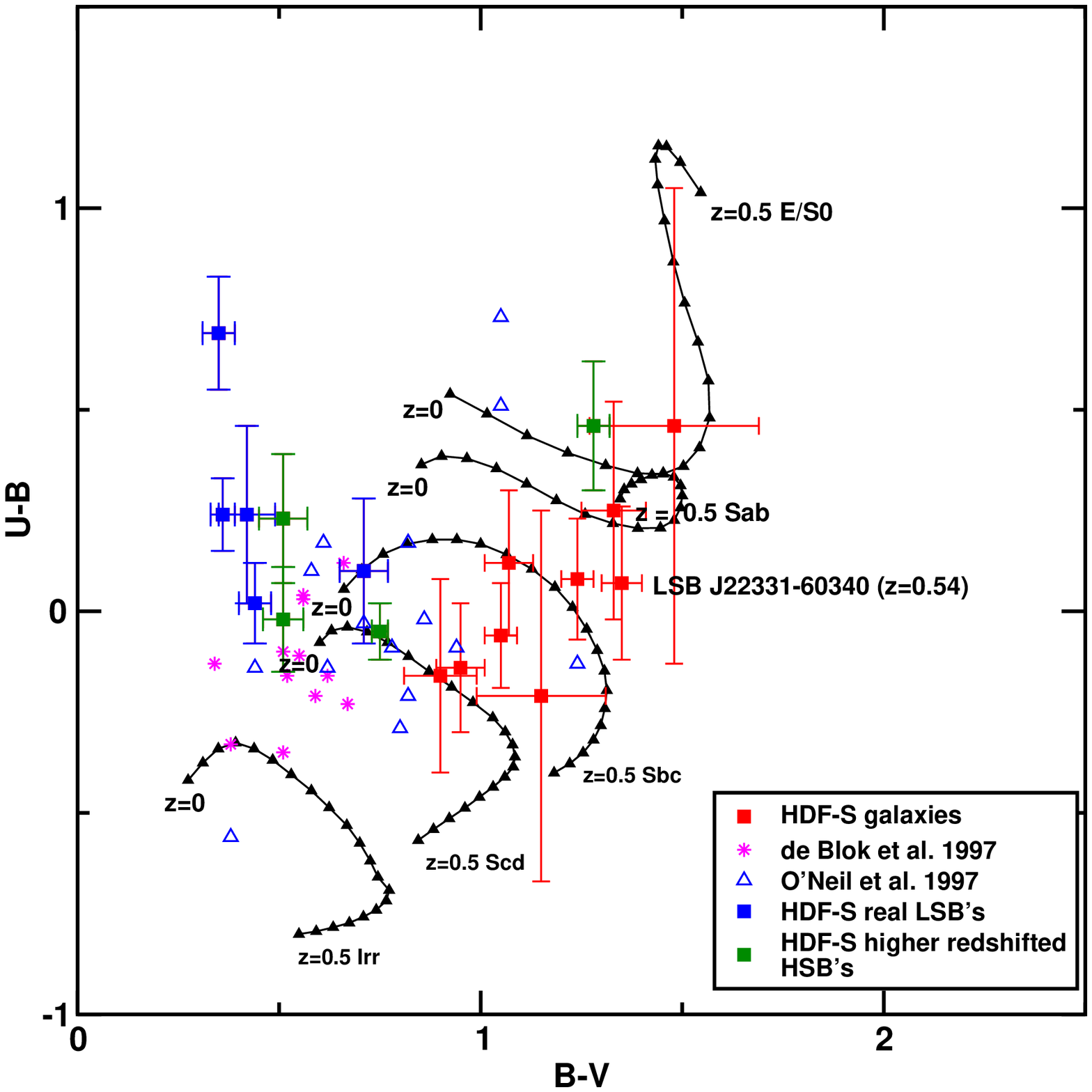}
\caption{Color-color diagrams $U\,-\,B$ vs. $B\,-\,R$ and
  $B\,-\,V$ vs. $U\,-\,B$ accounting for the spectroscopically derived distance
  information. From the selected LSB candidates (blue + green squares),
  56\,\% of the galaxies are found to be true LSBs (blue squares) and
  44\,\% are higher redshifted HSB galaxies (green squares). Two of the
  higher redshifted HSBs are located at z\,\sm\,0.1. They follow the redshift
  tracks of the ``normal'' HSB galaxies and therefore should not be selected
  as LSB candidates.}
\label{colorlsb}
\end{center}
\end{figure*}
In order to preselect LSB galaxies for spectroscopic follow-up
  observations efficient methods are required to distinguish candidates from,
  e.g., objects with low surface brightness caused by cosmological
  dimming. In a previous paper we have described such a method using a
  color-color selection criterion \citep{2007A&A...465...95H}.
  This selection criterion is
  based on the different location of the LSB candidates in comparison to the
  HSB redshift tracks in the color-color space (see also
  Fig.~\ref{colorlsb}).  With the spectroscopic distance information 
  we have now at hand for our LSB candidate sample we will discuss the
  efficiency of this method. 
 
  \quad
  Using this color-color selection we were able to clean our final sample from
  higher redshifted background sources (red squares in Fig.~\ref{colorlsb})
  which mimic LSB galaxies due to cosmological dimming effects, the so called
  ``Tolman-Dimming'' \citep{1935ApJ....82..302H}:
\begin{eqnarray}
\mu_{\nu_{0}}=\frac{\mu_{\nu_{1}}}{2.5log(1+z)^4}
\label{muz}
\end{eqnarray}
  For the selection process we used two different color-color diagrams
  (B\,-\,V vs. U\,-\,B and U\,-\,B vs. B\,-\,R) in combination with redshift
  tracks adopted from a work of \citet{1998AJ....116.1074L}. 

  After removing the higher redshifted background HSB galaxies from the
  candidate sample we were able to show, that most of the selected LSB galaxy
  candidates (7 galaxies, blue and green squares) have a distinct different
  location in the color-color diagrams compared to the redshift tracks of the
  five standard Hubble types (see Fig.~\ref{colorlsb}). 

  \quad 
  The LSB candidates have a redder U\,-\,B color
  (U\,-\,B\,\lareq\,-0.05\,mag), whereas the B\,-\,R and  B\,-\,V colors are
  shifted into the blue color range (B\,-\,R\loweq\,1.08\,mag,
  B\,-\,V\,\loweq\,0.71\,mag). 

  \quad
  Using the spectroscopically obtained distances we are now able to
  finally select the real LSB galaxies from our HDF-S LSB candidate sample (see
  Table~\ref{lsbsamp}). It turned out that \si\,45\% of the preselected
  LSB candidates (4 out of 9) have redshifts z\,\loweq\,0.1 (blue squares in
  Fig.~\ref{colorlsb}). These galaxies are local galaxies. No surface
  brightness correction against cosmological dimming has to be applied and,
  therefore, they are real LSB galaxies. One of the candidates
  (\object{LSB J22352-60420}) has a slightly larger redshift of
  z\,=\,0.12. However, after correcting the central surface brightness against
  ``Tolman-Dimming'', this galaxy still remains as a LSB galaxy, showing a
  central surface brightness below \m$_0$\,$>$\,22.5\,$B$\,\magarc\, (see
  Table~\ref{lsbsamp}). Finally the majority (5 out of 9) of the
  preselected LSB candidates turned out to be genuine LSB galaxies. For our
  final LSB galaxy sample we include all galaxies having central surface
  brightnesses $\mu_0\,\ge\,22.2$\,\magarc, which is more than 2\sig\,
  below the  Freeman value of $\mu_0\,=\,21.65\pm\,0.3$\,\magarc
  \citep{1970ApJ...160..811F}. 
\begin{table*}
\centering
\begin{tabular}{c c c c c c c}
\hline
\hline
Name&U-B&B-V&B-R&\m$_0$&\m$_0$(corr)&LSB galaxy\\
&[mag]&[mag]&[mag]&$B_\mathrm{W}$\magarc&$B_\mathrm{W}$\magarc&\\
\hline
\object{LSB J22311-60503}&-0.05&0.75&1.21&22.4&21.9&no\\
\object{LSB J22324-60520}&0.46&1.28&1.99&22.3&21.6&no\\
\object{LSB J22325-60155}&0.02&0.44&0.66&23.3&23.3&yes\\
\object{LSB J22330-60543}&0.24&0.36&0.55&22.2&22.2&yes\\
\object{LSB J22343-60222}&0.24&0.42&0.64&23.3&23.3&yes\\
\object{LSB J22352-60420}&0.10&0.71&1.08&23.2&22.7&yes\\
\object{LSB J22353-60311}&0.69&0.35&0.77&23.4&23.4&yes\\
\object{LSB J22354-60122}&-0.02&0.51&0.78&22.5&21.8&no\\
\object{LSB J22355-60183}&0.23&0.51&0.79&22.5&22.0&no\\
\hline
\end{tabular}
\caption{List of the galaxies which were selected as LSB
  candidates using color-color diagrams. The last column marks whether the
  candidate, after analyzing the spectroscopic information, is a true LSB
  galaxy or not. The central surface brightness in column 6 is corrected
  against fading due to ``Tolman-Dimming''.}
\label{lsbsamp}
\end{table*} 
  The other four selected candidates (green squares) show redshifts
  z\,\lareq\,0.1. After correcting against the ``Tolman-Dimming'' effect, the
  central surface brightnesses of these candidates came out to have values
  \m$_0$\,$\le$\,22.2\,\magarc. Therefore, these galaxies have to be
  considered as higher redshifted ``normal'' HSB galaxies.

  \quad
  After a first inspection, it turned out that by using the color-color
  selection criterion, only 56\,\%  of the selected galaxies are genuine LSB
  galaxies. However, this selection criterion also resulted in nearly the same
  amount (44\,\%) of higher redshifted ``normal'' HSB background galaxies. 
  Therefore, this would not be a very reliable method in order to preselect a
  LSB subsample from a large sample of galaxies. A more carefully analysis of
  the selection in Fig.~\ref{colorlsb} shows that two of the higher redshifted
  HSB galaxies (\object{LSB J22311-60503}, \object{LSB J22324-60520}, green
  squares) which were also selected as LSB candidates do follow the redshift
  tracks. After correcting the surface brightnesses against
  ``Tolman-Dimming'', these galaxies appear to be higher redshifted ``normal''
  HSB galaxies. Their classification as LSB candidates was too optimistic, due
  to the relatively large uncertainties in the photometric redshift
  determination. Two additional galaxies (\object{LSB
  J22354-60122},\object{LSB J22355-60183}, green squares) which are clearly
  separated from the redshift tracks also turned out to be located at
  redshifts z\,\lareq\,0.1. After correcting their central surface
  brightnesses against ``Tolman-Dimming'', these galaxies also 
  came out to be HSB galaxies. However, these two galaxies do not behave like
  ``normal'' HSBs. They must have extreme colors in order to be moved to this
  location in the color-color space.

  \quad  
  Excluding now the two higher redshifted galaxies, located along the redshift
  tracks, the method yields 5 LSB galaxies (71\,\%) out of a sample of 7
  selected galaxies. Only 2 out of 7 galaxies (29\,\%) are higher redshifted
  HSB galaxies. This relatively high success rate makes the described
  method a reasonable tool to select LSB galaxies from larger samples of
  galaxies.

\subsection{Volume densities of the HDF-S LSB galaxies}

  One result of the search for LSB galaxies in the HDF-S is a number surface 
  density of 8.5 LSB galaxies per deg\2. This is more than two times higher
  than found in previous surveys \citep[e.g.,][which give 4 LSB galaxies
  deg$^{-2}$]{1997AJ....113.1212O,1997AJ....114.2448O,1997AJ....114..635D}.
  This number density is not very meaningful, since due to the use of much more
  sensitive data ($\mu_{lim}\,\approx\,$29 \magarc), a much larger search
  volume was covered. 

  \quad 
  In order to also include the covered volume V, it is more convenient to
  calculate the volume density $n$ for the derived sample. This we have done
  following a method from \citet{1996MNRAS.280..337M}. With this method,
  surface brightness corrected and normalized volume densities \Ph\, could be
  estimated. At this point we like to mention that we are aware that the
  method described in the following is a statistical method to compare the
  surface brightness distribution of different galaxy samples. Our sample
  of five LSB galaxies plus three extreme LSB candidates (description see
  below) is 
  not a statistically significant galaxy sample. Thus,
  any interpretations drawn from this analysis have to be discussed very
  carefully. However, using this kind of analysis give hints about how our
  results fit into the existing overall picture of LSB galaxies.

  \quad  
  The relative volume density of LSB galaxies $n_{LSB}$ in relation to the
  volume density of a well known sample of HSB galaxies $n_{HSB}$ is
  described as: 
\begin{eqnarray}
\frac{n_\mathrm{LSB}}{n_\mathrm{HSB}}&=&\frac{N_\mathrm{LSB}}{N_\mathrm{HSB}}\left(\frac{(V/\Omega)_\mathrm{HSB}}{(V/\Omega)_\mathrm{LSB}}\right)\\
\Phi_\mathrm{norm}&=&\frac{n_\mathrm{LSB}}{n_\mathrm{HSB}}=\frac{N_\mathrm{L}}{N_\mathrm{H}}\left(\frac{\theta_\mathrm{l}^\mathrm{L}}{\theta_\mathrm{l}^\mathrm{H}}\right)^3\left(\frac{\mathrm{h}_\mathrm{H}}{\mathrm{h}_\mathrm{L}}\right)^3\frac{(\mu_\mathrm{l}^\mathrm{H}-\mu_0)^3}{(\mu_\mathrm{l}^\mathrm{L}-\mu_0^\mathrm{L})^3}
\label{phinorm}
\end{eqnarray} 
  Most of the parameters in Eqn.~\ref{phinorm} can be derived without having
  distance information for the sample galaxies. The only exception is the
  scale-length h in physical units. For the calculation of the volume density,
  the assumption is made that the scale-length $h_L$ of the derived LSB galaxy
  sample is comparable to the scale-length $h_H$ of the sample of HSB galaxies
  to which the sample will be normalized ($h_H/h_L\,=\,1$). If, on average,
  HSB galaxies had a larger scale-length, the value of $n_{LSB}/n_{HSB}$
  would increase. Using this assumption, we were able to calculate lower
  limits for the normalized volume densities of our LSB galaxy sample.

  \quad
  For our HDF-S LSB sample, we determined a number surface density of
  $N_L$\,=\,8.5 deg$^{-2}$ as well as a surface brightness limit of
  $\mu_l$\,=\,29.0\,\magarc. We chose a diameter limit of
  $\theta_l$\,$\ge$\,10.8\,arcsec. Due to the small number of objects, we used
  a central surface brightness bin size of 1\,\magarc\, in order to estimate
  normalized volume densities. For the first two surface brightness bins
  (22.5\,\magarc\, and 23.5\,\magarc) we calculated the volume densities using
  a field size of 0.59\,deg$^2$. Therefore, we accounted for the fact that all
  of these galaxies were found in the smaller field covered by the Goddard
  data \citep[see][ for more details]{2007A&A...465...95H}. For the extreme
  LSB candidates (last two surface brightness bins) we had to choose the larger
  NOAO field of 0.76\,deg$^2$. Finally we normalized the sample with
  respect to the well studied galaxy sample of
  \citet{1996A&AS..118..557D}. This is 
  a statistically complete sample of 86 disk galaxies. The sample was selected
  from the UGC catalog including galaxies with diameters $\theta$ larger
  than 2\,arcmin and a minor to major axis ratio larger than 0.625. The survey
  covered an area of 1.57\,srad\,$\equiv$\,5154.3\,deg$^{-2}$. The surface
  brightness limit of the catalog is about $\mu_l$\,=\,26.5\,$B$\,\magarc. We
  calculated the errors for the surface brightness corrected volume densities
  using Poisson statistics and Gaussian errors perturbation:
\begin{eqnarray}
\sigma_\mathrm{L}&=&\sqrt{\frac{1}{\cal N_{L}}}\\
\Delta\Phi_\mathrm{norm}&=&\pm\sqrt{\left(\frac{\partial\Phi_\mathrm{norm}}{\partial
      N_\mathrm{L}}\,\sigma_\mathrm{L} \right)^2\,+\,\left(\frac{\partial\Phi}{\partial\mu_0^\mathrm{H}}\Delta\mu_0^\mathrm{H}\right)^2}   
\end{eqnarray}     
  For comparison we also plotted the surface brightness corrected volume
  densities (estimated following the same method) for two additional, well
  known LSB samples of \citet{1997AJ....113.1212O} (Texas Survey) and
  \citet{1996ApJS..105..209I}. The Texas survey covered an area of 27\,deg\2\,
  on the sky. In this area, a sample of 127 LSB galaxies with diameters larger
  than 13.5\,arcsec was derived. The survey reached a surface brightness limit
  of $\mu_l$\,=\,27\,$B$\,\magarc. The search area of the survey of
  \cite{1996ApJS..105..209I} covers about 768 deg\2\ of the sky. The field is
  located around the equatorial strip over a declination range of
  $\pm$\,2$^{\circ}$. The survey reached a surface brightness limit of
  $\mu_l$\,=\,26\,$B$\,\magarc. The derived sample consists of 693 galaxies
  larger than 30\,arcsec. For both surveys we used a surface brightness bin
  size of 0.5\,\magarc. 
\begin{figure}[ht]
\begin{center}
\includegraphics[width=8cm]{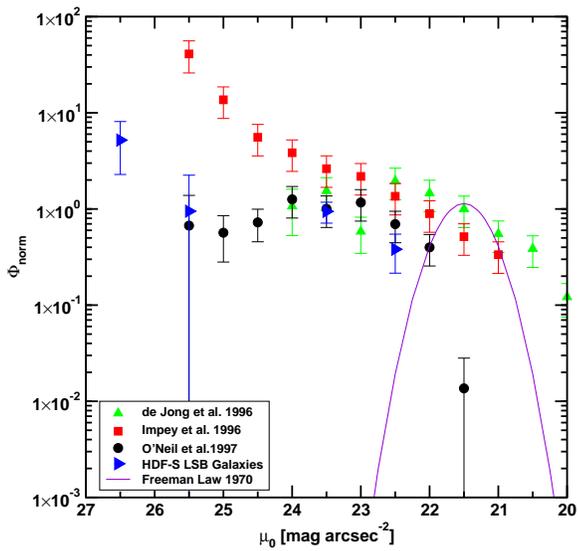}
\caption{Surface brightness corrected volume densities
  for several samples of LSB galaxies and the HSB galaxy sample of
  \citet[][green triangle]{1996A&AS..118..557D}. The LSB galaxy sample
  derived in the HDF-S is represented by the blue triangles. For a comparison,
  the distribution of the galaxies expected from the Freeman Law
  \citep[][purple line]{1970ApJ...160..811F} is plotted. The diagram shows
  that in contrast to the expected distribution of the Freeman Law, the
  distribution for the LSB galaxies stays flat down to very low central
  surface brightnesses.}
\label{volumedens}
\end{center}
\end{figure}

  \quad
  The results of the discussed estimations of the normalized, surface
  brightness corrected volume densities is plotted in
  Fig.~\ref{volumedens}. In this diagram, we were able to show that the
  central surface brightness distribution for our HDF-S LSB galaxy sample
  (LSB J's, blue triangles) follows the flat distribution of the other LSB
  surveys. At lower central surface brightnesses (more than 3 \sig\, below the
  Freeman value of \m$_0$\,=\,21.65\,\plmi0.35\,\magarc ), the number of LSB
  galaxies is much higher than expected from the Freeman Law (purple
  line). 
  The volume density of the HDF-S LSB galaxy sample derived
  for the surface brightness bin \m\,=22.5 \magarc\, is slightly lower
  compared to the values of
  \citet[][]{1997AJ....113.1212O,1996ApJS..105..209I} and
  \cite{1996A&AS..118..557D}. This is maybe the result of incompleteness at
  the upper surface brightness limit of the HDF-S galaxy sample (upper
  selection limit \m(0)\,=\,22.0\magarc) and/or due to very low number
  statistics for the HDF-S LSB sample. The Texas survey also shows an
  incompleteness around the Freeman value (\m\,=\,21.5\,\magarc). The large
  error bar of our survey at \m\,=\,25.5\,\magarc\,resulted from low number
  statistic for this bin (1 LSB galaxy). However, the one object at this value
  and the two 
  objects at the next bin (\m\,=\,26.5\,\magarc) still have a big implication
  for the volume density, due to the low volume over which such extreme Low
  Surface Brightness galaxies could be detected. To find one of these extreme
  LSB galaxies in a small field, as covered by the NOAO data
  ($\sim$\,0.76\,deg$^2$), has a very low probability. The detection of these
  objects indicate that the formation of such extreme LSB galaxies is as
  common in the Universe as the formation of the higher surface brightness
  objects found in the surveys.  
 
\subsection{Metallicities}

\begin{figure}[ht]
\begin{center}
\includegraphics[width=8cm]{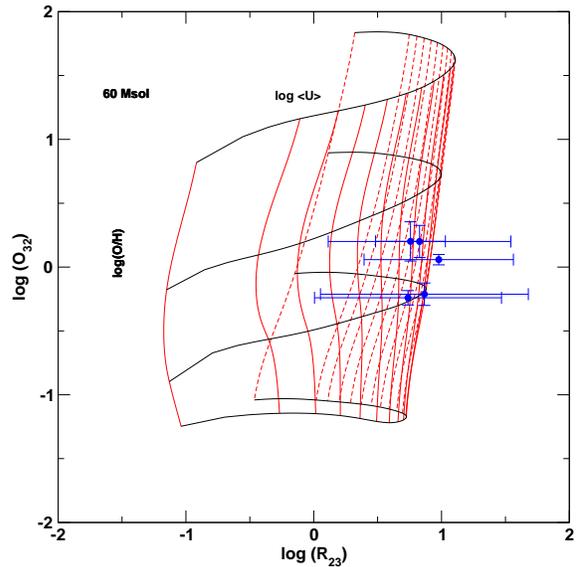}
\caption{Metallicities for the five LSB galaxies in the HDF-S, derived using
  the strong-line method based on the semi-empirical calibration described in 
  \citet{1994ApJ...426..135M} and \citet{2004MNRAS.355..887D}. }
\label{metallicities}
\end{center}
\end{figure}

\begin{table*}
\centering
\begin{tabular}{c c c c c c c c c c c}
\hline
\hline
Name&H$\beta$&[O{\sc ii}]&[O{\sc iii}]$\lambda$4959&[O{\sc
  iii}]$\lambda$5007&[N{\sc ii}]&log(R23)&log(O32)&log([N{\sc ii}]/[O{\sc ii}])&log(O/H)& X$\cdot$Z$_\odot$\\
&[\AA]&[\AA]&[\AA]&[\AA]&[\AA]&&&&&\\
\hline
\object{LSB J22325-60155}&4.42\plmi1.00&11.50\plmi1.50&5.35\plmi1.67&12.9\plmi1.60&0.80&0.83\plmi0.71&0.20\plmi0.13&-1.16&-3.96\plmi1.30&0.13\\
\object{LSB J22330-60543}&0.96\plmi0.13&4.26\plmi0.26&1.35\plmi0.19&3.52\plmi0.19&0.25&0.98\plmi0.58&0.06\plmi0.04&-1.23&-3.63\plmi1.15&0.29\\
\object{LSB J22343-60222}&1.79&3.95\plmi0.67&1.95\plmi0.65&4.33\plmi0.63&0.26&0.76\plmi0.27&0.20\plmi0.15&-1.18&-4.09\plmi0.48&0.10\\
\object{LSB J22352-60420}&0.67\plmi0.14&3.05\plmi0.47&0.63\plmi0.39&1.24\plmi0.38&1.11&0.87\plmi0.81&-0.21\plmi0.09&-0.44&-3.54\plmi1.04&0.37\\
\object{LSB J22353-60311}&1.70\plmi0.49&5.90\plmi0.81&1.01\plmi0.32&2.38\plmi0.52&0.91&0.74\plmi0.73&-0.24\plmi0.06&-0.81&-3.37\plmi0.74&0.54\\
\hline
\end{tabular}
\caption{ Results of the strong-line method, derived using the
    semi-empirical calibration described in \citet{1994ApJ...426..135M} and
  \citet{2004MNRAS.355..887D}. Columns 2 to 6 show the measured equivalent
  width (including errors) for the LSB galaxy sample). Columns 7 to 9 gives
  the logarithm of the emission line relations as used in the R$_{23}$
  method, column 10 gives the oxygen abundances and the last column shows the
  oxygen abundance in terms of solar abundance. The values of log([N{\sc
  ii}]/[O{\sc ii}]) are derived using upper limits for the [N{\sc ii}]
  measurements. The metallicities are derived from the lower branch values for
  log([N{\sc ii}]/[O{\sc ii}])$<$-1 and from the upper branch for log([N{\sc
  ii}]/[O{\sc ii}])$>$ -1.}
\label{lsbmetall}
\end{table*}
Since we were not able to detect emission lines sensitive for direct abundance
measures (e.g O[{\sc iii}]$\lambda$4363), we have to use a different
approach to get information about the metallicities in our LSB galaxy sample. 
One commonly used method to derive abundance information is the strong-line
method of \citet{1979MNRAS.189...95P}. We used this so called R23-method in
combination with the semi-empirical calibration of
\citet{1991ApJ...380..140M,1994ApJ...426..135M} to get an estimation of the 
oxygen abundances in the derived HDF-S LSB galaxies. With the R23-method we
were able to estimate metallicities just using measurements of the most
prominent emission lines H\B, [O{\sc ii}], and the [O{\sc iii}]-doublet (see
Table~\ref{lsbmetall}).
\begin{eqnarray}
R_{23}\,&=&\,\frac{[OII]\lambda 3727+[OIII]\lambda\lambda4959,5007}{H\beta}\\
\nonumber\\
O_{32}\,&=&\,\frac{[OIII]\lambda\lambda4959,5007}{[OII]\lambda 3727}
\end{eqnarray}
The semi-empirical calibration which we used for this abundance estimation is
described in more detail in Appendix~A of \citet{2004MNRAS.355..887D}. To
discriminate between the upper and lower branch of this solution (see
Fig~\ref{metallicities}) we used the [N{\sc ii}] over [O{\sc ii}] relation as
described in \citet{1994ApJ...426..135M}. Objects with
$log([NII]/[OII])\,>\,-1$ had abundances derived from the upper
brunch, while for $log([NII]/[OII])\,<\,-1$ we used the solution from
the lower branch. 

Despite of the large uncertainties for the derived emission line intensities  
we derived subsolar oxygen abundances for all LSB galaxies in our
sample (see Table~\ref{lsbmetall}).
The oxygen abundance range between $\sim$ 1/2 Z$_\odot$ and $\sim$
1/10th Z$_\odot$. 
These low metallicities are typical for LSB galaxies and in good agreement
with other measurements
\citep[e.g.][]{1994ApJ...426..135M,1998A&A...335..421D,2004MNRAS.355..887D}.

\section{Conclusions and Summary}

  We presented results of spectroscopic follow up observations, obtained
  using the ESO 3.6\,m telescope, of a sample
  of 9 LSB galaxy candidates, derived in a 0.59\,deg$^2$ field around the
  HDF-S. We used the measured emission lines to determine distances for
  the observed galaxies in the redshift range of z\,\si\,0.05 to
  z\,\si\,0.16. The majority  of the observed HDF-S sample galaxies is located
  at redshifts below z\,\loweq\,0.1, which makes them real LSB galaxies. Using
  color-color diagrams ($B\,-\,V$ vs. $U\,-\,B$ and $U\,-\,B$ vs. $B\,-\,R$),
  we could  show that 
  71\,\% of the galaxies, having a significant different location compared to
  the location of the HSB galaxies, are genuine LSB galaxies. This indicates
  that the use of the location in the color-color diagrams is an efficient
  method in order to preselect the content of LSB candidates against higher
  redshifted background galaxies. In the color-color diagrams, the LSB
  galaxies are shifted to the blue region for the $B-R$ and $B-V$ color and to
  the 
  red region for the U-B color. This different location in the color-color
  space seems to be a hint for a different stellar population mix, and
  therefore, a different Star Formation History (SFH) of the selected LSB
  galaxies. The shifts could be caused by a stronger Balmer jump indicating
  younger stellar population. The shift into the red region of the $U-B$ color
  could result from a low UV flux due to a recent low formation rate of O, B,
  and A stars. The shift to the blue of the $B-R$ and $B-V$ color also
  seems to indicate that there exists no strong underlying old stellar
  population which could cause the shift to the red in the $U-B$
  color. To produce this blue-ward shift, a steeper decline of the red
  part of the stellar continuum emission is needed. Therefore, the red U-B
  color could also not be
  the result of a higher metalicity. In this case, the decline in the red
  part of the continuum emission must be shallower, caused by a stronger
  underlying old stellar population. The suppression of the $U$-flux due
  to a large amount of dust could also be excluded. Until now only a few
  detections of LSB galaxies in the FIR wavelength region at very low
  levels exist \citep{1994AJ....108..446H}.  

 \quad
  We were also able to estimate oxygen abundances of our five LSB galaxies in
  the HDF-S, using the strong-line method in combination with the
  semi-empirical calibration of \citet{1991ApJ...380..140M}. Despite the large
  uncertainties, we estimated relatively low oxygen abundances between 0.5 to
  0.1 of solar value. This indicates that the sample LSB galaxies are
  relatively young and unevolved and could explain their location in the
  color-color space.

  \quad
  The derived LSB sample consists of galaxies with scale-length between
  2.5-7.3~kpc and absolute $B$-band magnitudes between $M_B$\,=\,-16.90 mag
  and $M_B$\,=\,-18.67 mag (see Table~\ref{physpar}). From the absolute
  magnitudes, luminosities were derived in the range of
  0.51\,-\,2.61\,$\cdot$\,10$^9$L$_{\odot}$. Therefore, the sample does not
  include dwarf galaxies, which is often expected for LSB galaxy samples. All
  galaxies have distances larger than z = 0.05. No nearby LSB galaxy could be
  found, perhaps not surprising given our small survey volume at very low
  redshift.   

  \quad
  In recent years, more sensitive surveys have shown that the Freeman Law
  \citep{1970ApJ...160..811F} was the result of selection effects (see
  Fig.~\ref{volumedens}). From these surveys, surface brightness distributions
  which stay flat down to central surface brightnesses of about
  25.5\,\magarc\, could be obtained. An estimation of the surface brightness
  dependent volume density for the HDF-S LSB sample indicate that the results
  of the HDF-S LSB survey are consistent with results found for other surveys
  \citep[e.g.,][]{1997AJ....113.1212O,1997AJ....114.2448O,1996ApJS..105..209I,1996A&AS..118..557D}.
  Additionally, the distribution could be expanded down to very low surface
  brightnesses of \m$_0$\,=\,27\,\magarc. It could be shown that the
  distribution stays flat also for these very low central surface
  brightnesses. The results of the HDF-S  LSB sample fit very well to the
  picture drawn by former surveys and lead to the assumption that the LSB
  galaxies represent a major contribution of the local galaxy
  population. \citet{astro-ph0006253} stated that if the distribution stays
  flat down to a central surface brightness around 30\,\magarc, up to 50\,\%
  of the local galaxy population could consist of LSB
  galaxies. \citet{2004MNRAS.355.1303M} showed that 60\,\% of the population
  of gas rich disks is represented by the LSB galaxies. All together the LSB
  galaxies represent a non-neglectable part of the local galaxy population,
  and therefore, they play an important role for the understanding of the
  formation and evolution processes of galaxies in general. 

\begin{figure*}[ht]
\centering
\includegraphics[width=7.5cm]{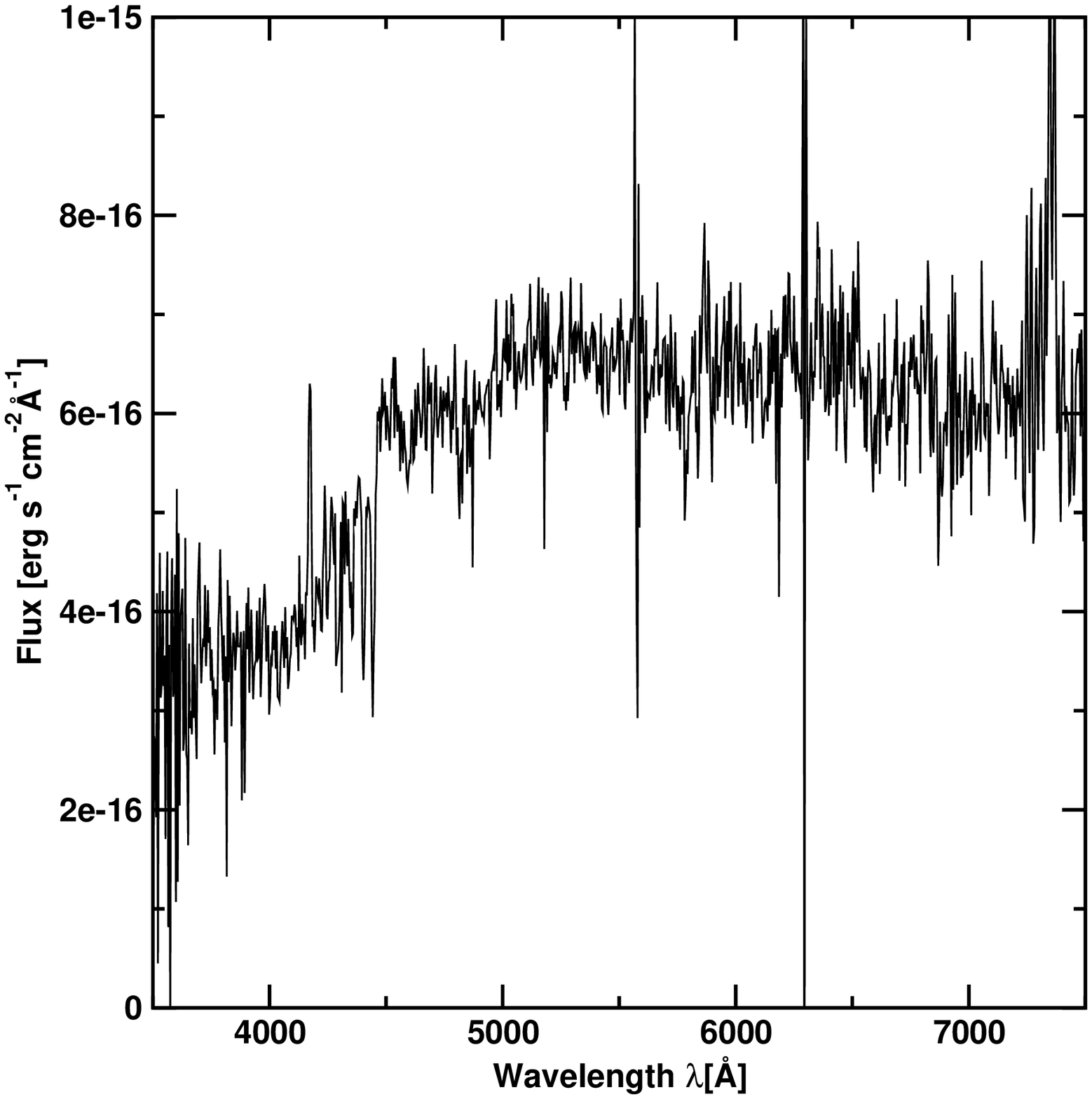}
\includegraphics[width=8cm]{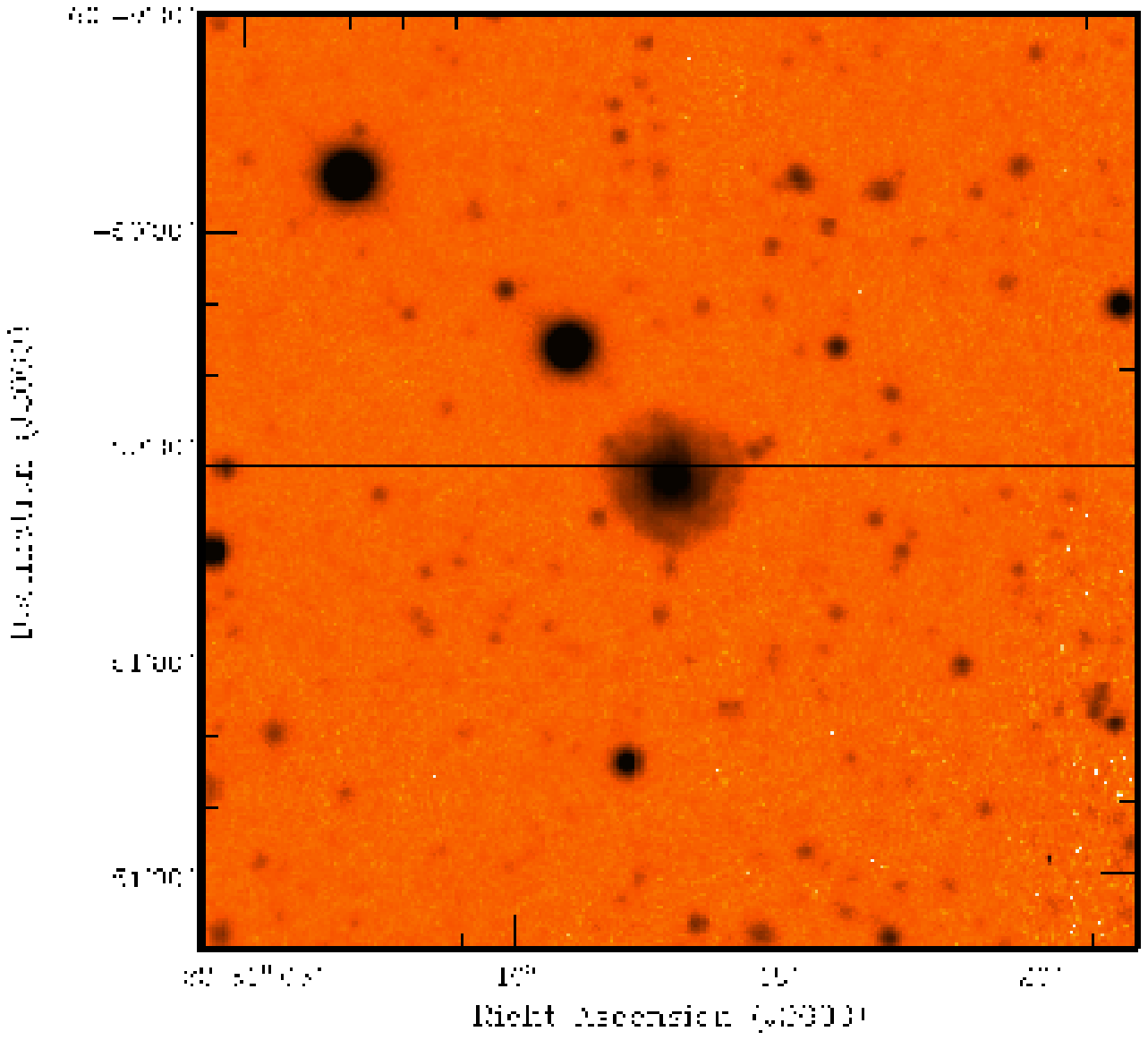}
\includegraphics[width=7.5cm]{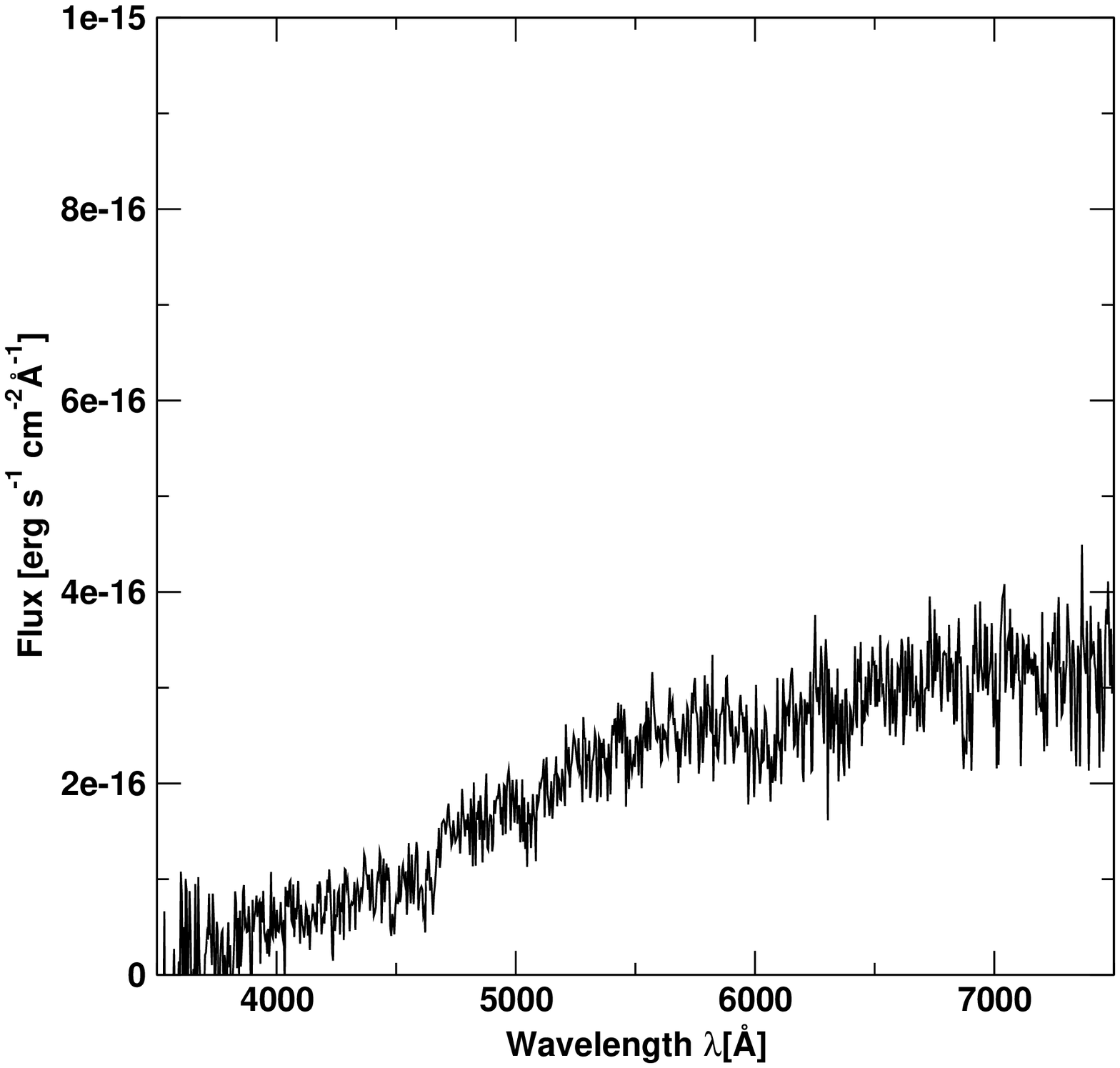}
\includegraphics[width=8cm]{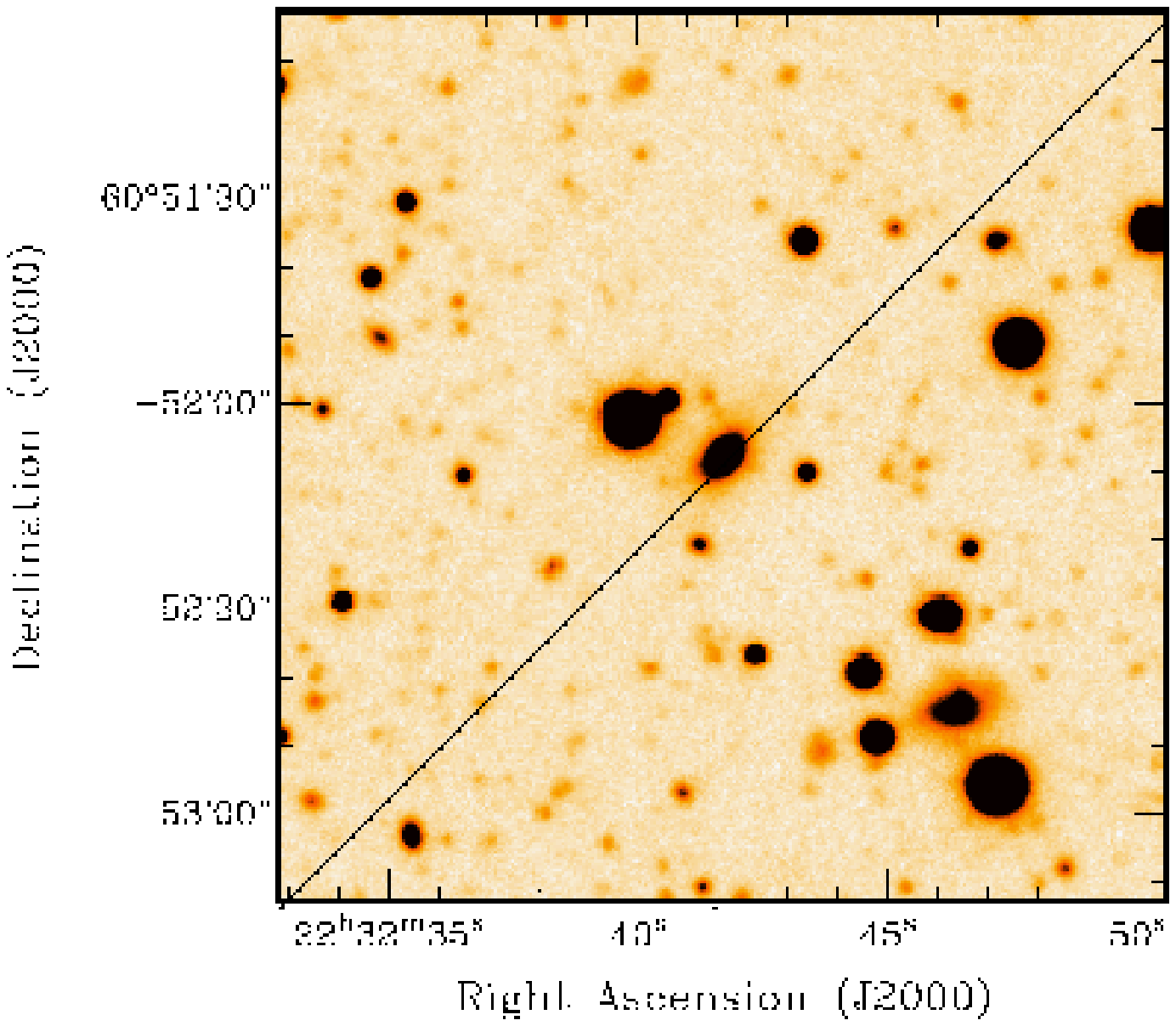}
\parbox{18cm}{{\bf Fig. 1 continued:} From top to bottom:
  \object{LSB J22311-60503} and \object{LSB J22324-60520}.}
\end{figure*}

\begin{figure*}[ht]
\centering
\includegraphics[width=7.5cm]{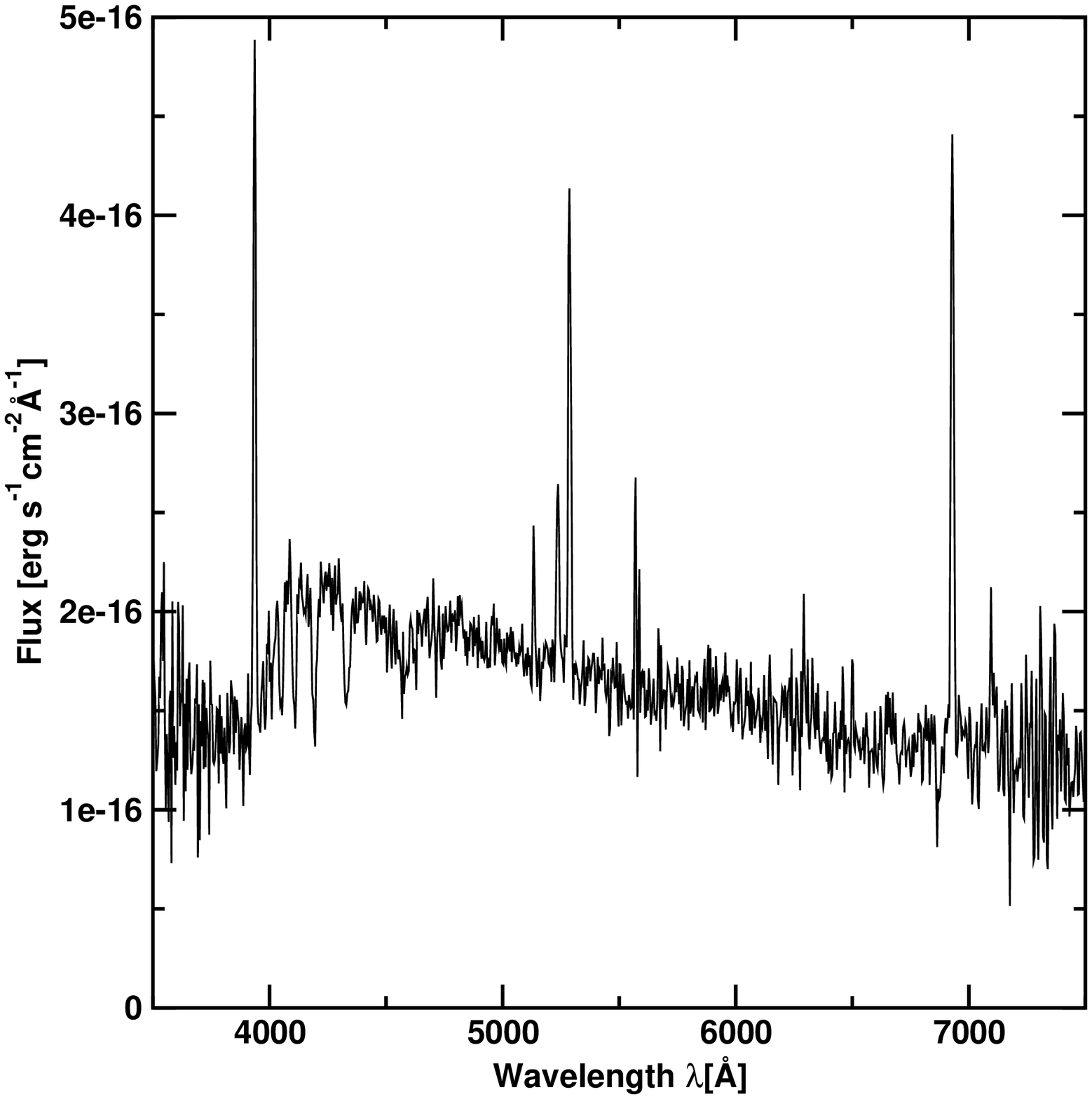}
\includegraphics[width=8cm]{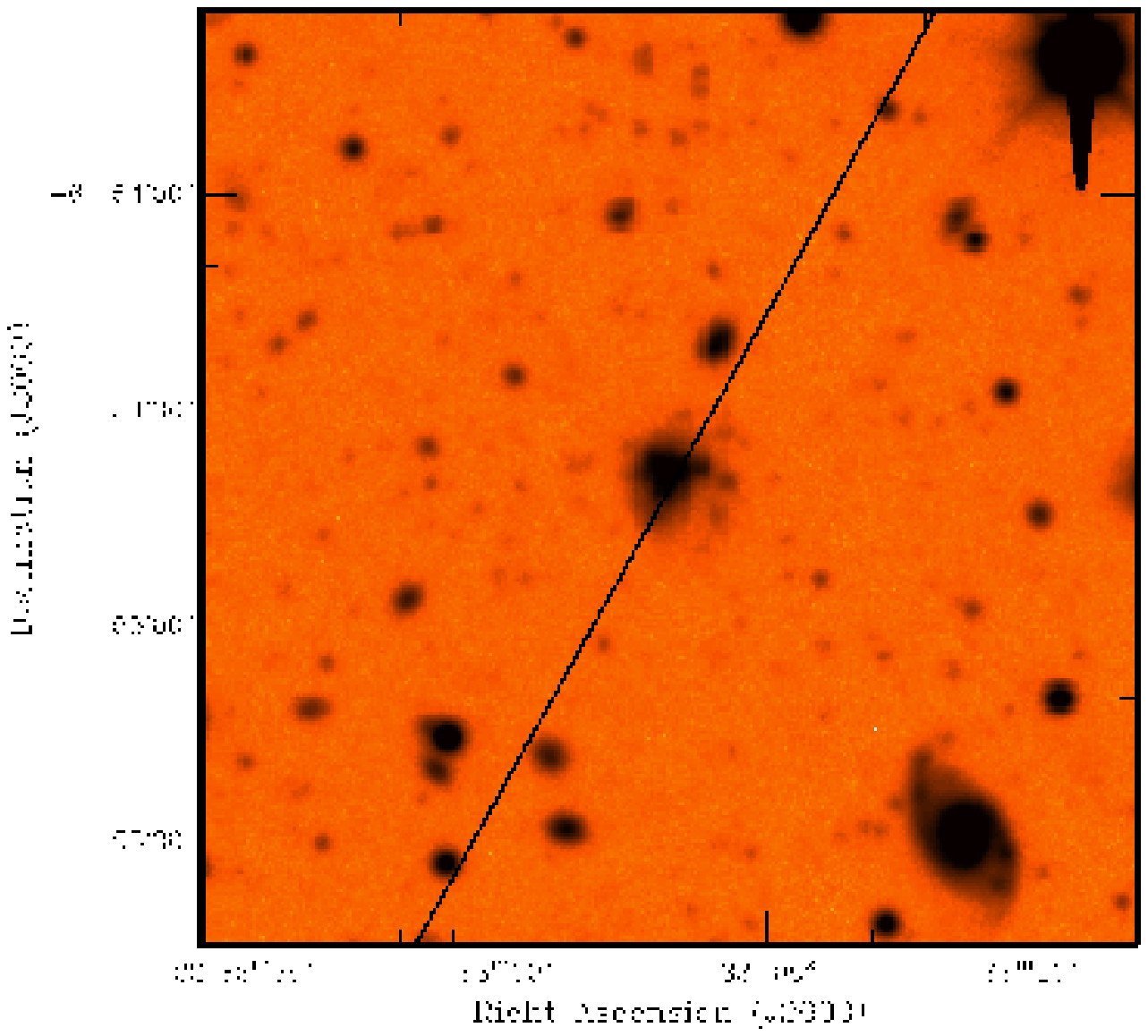}
\includegraphics[width=7.5cm]{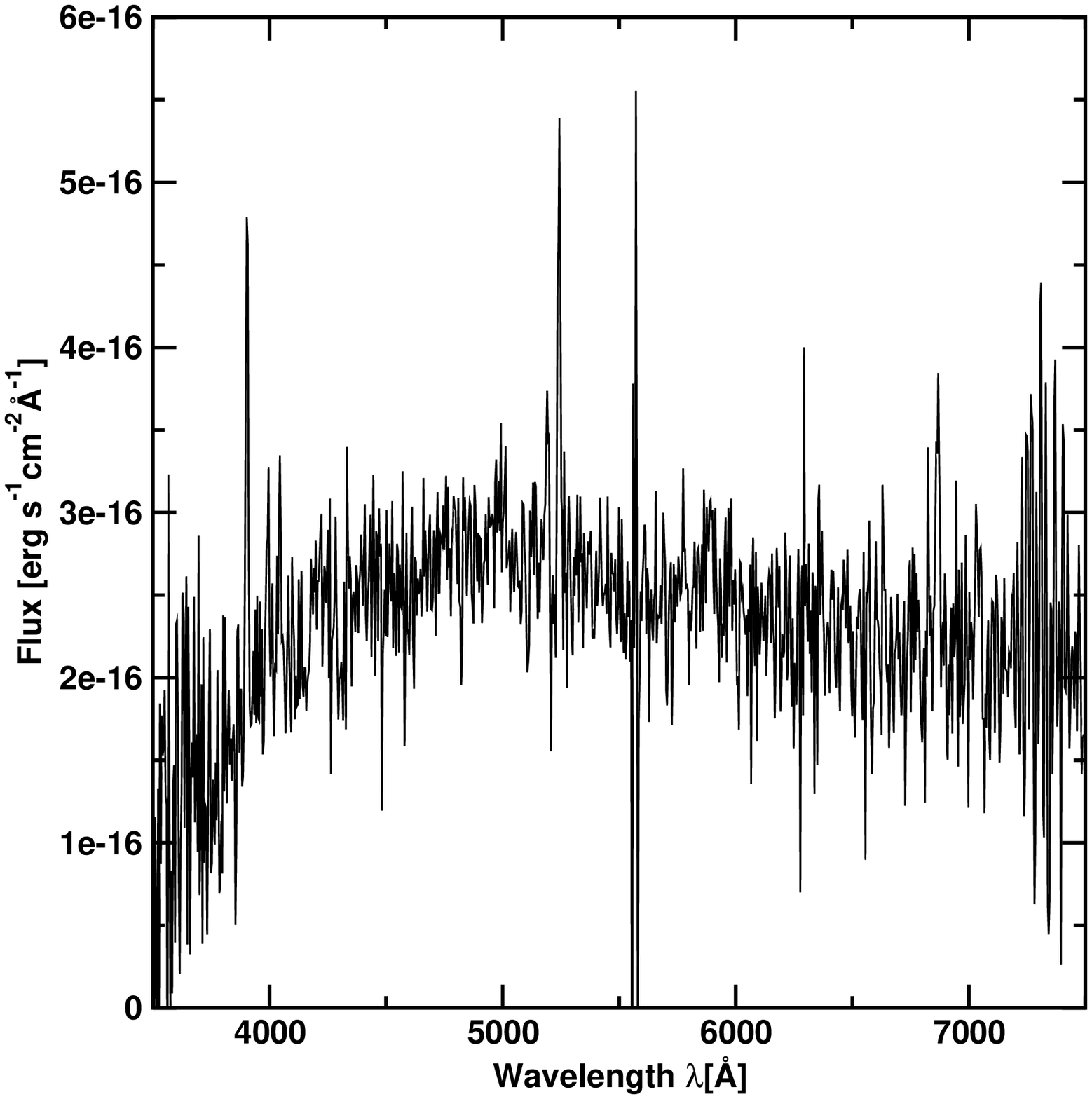}
\includegraphics[width=8cm]{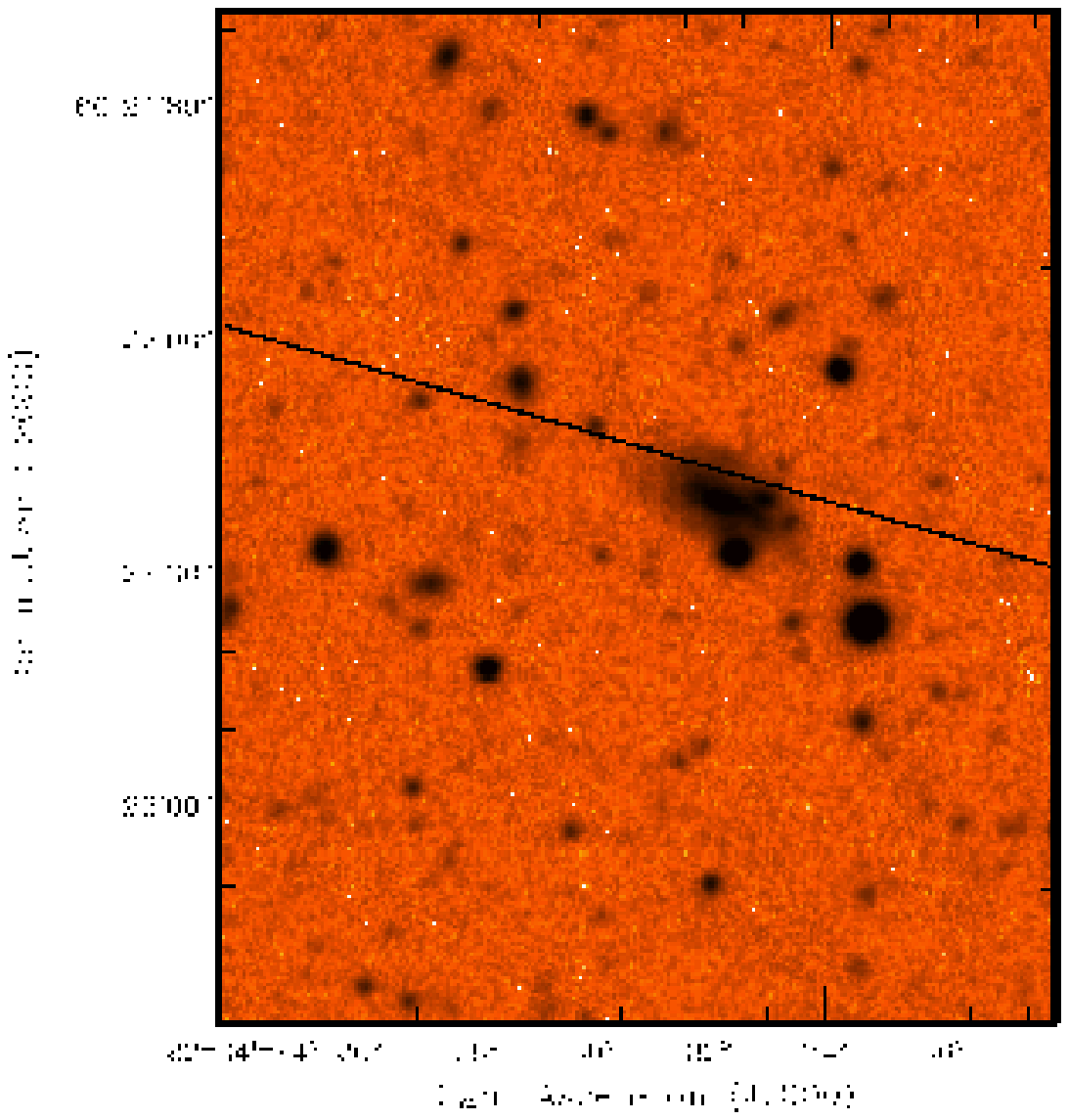}
\includegraphics[width=7.5cm]{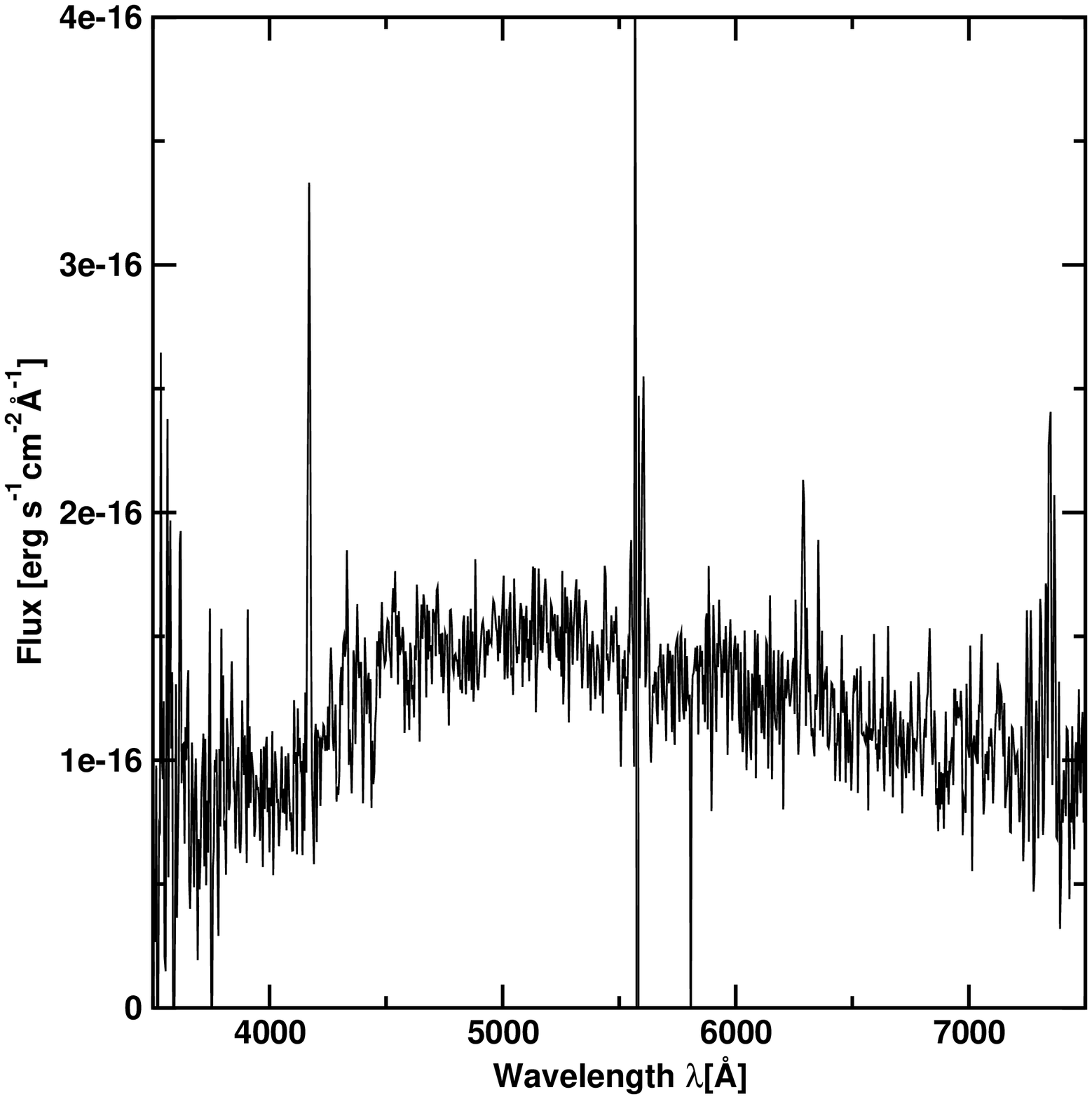}
\includegraphics[width=8cm]{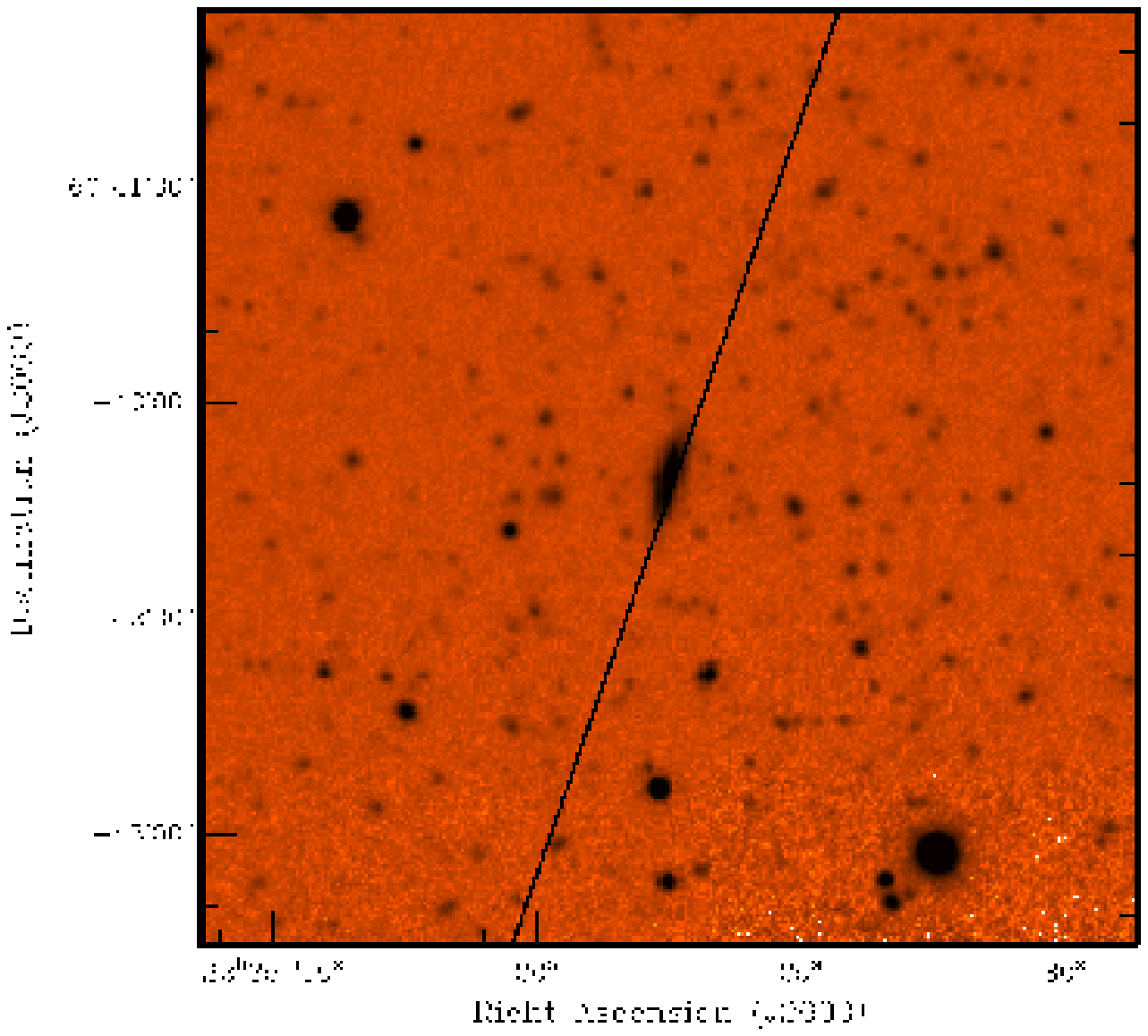}
\parbox{18cm}{{\bf Fig. 1 continued:} From top to bottom:
  \object{LSB J22330-60543}, \object{LSB J22343-60222}, \object{LSB J22352-60420}}
\end{figure*}
\begin{figure*}[ht]
\centering
\includegraphics[width=7.5cm]{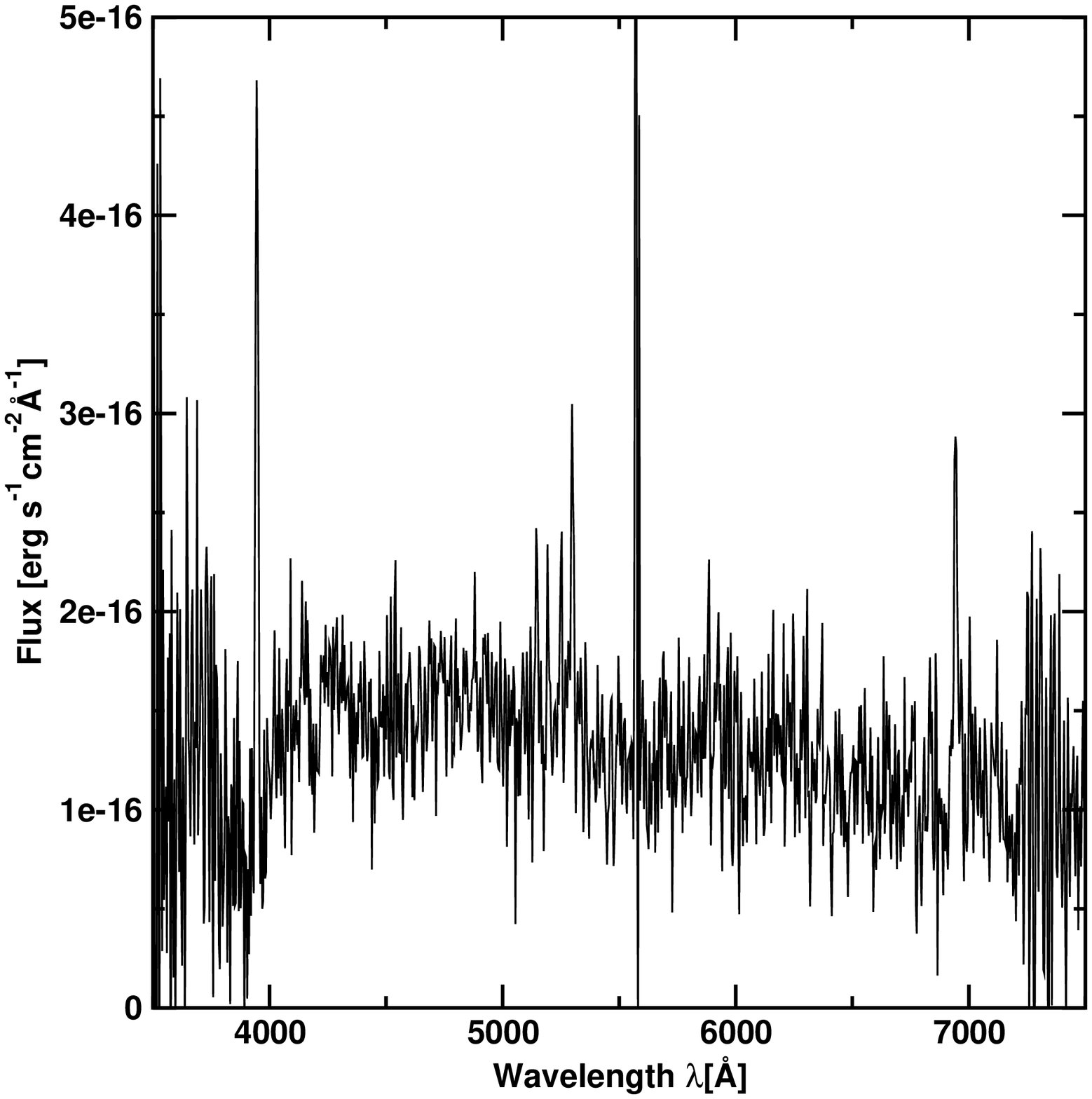}
\includegraphics[width=8cm]{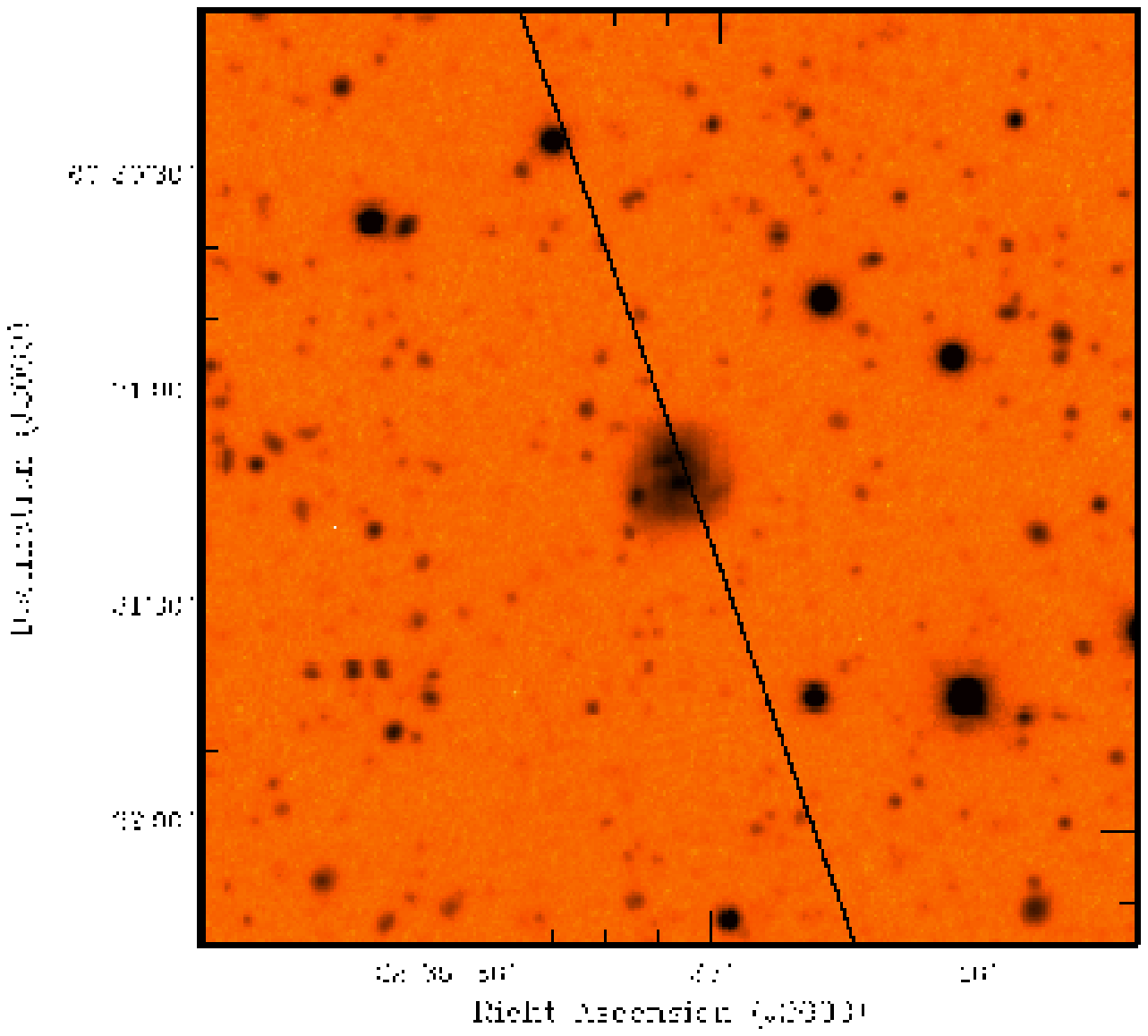}
\includegraphics[width=7.5cm]{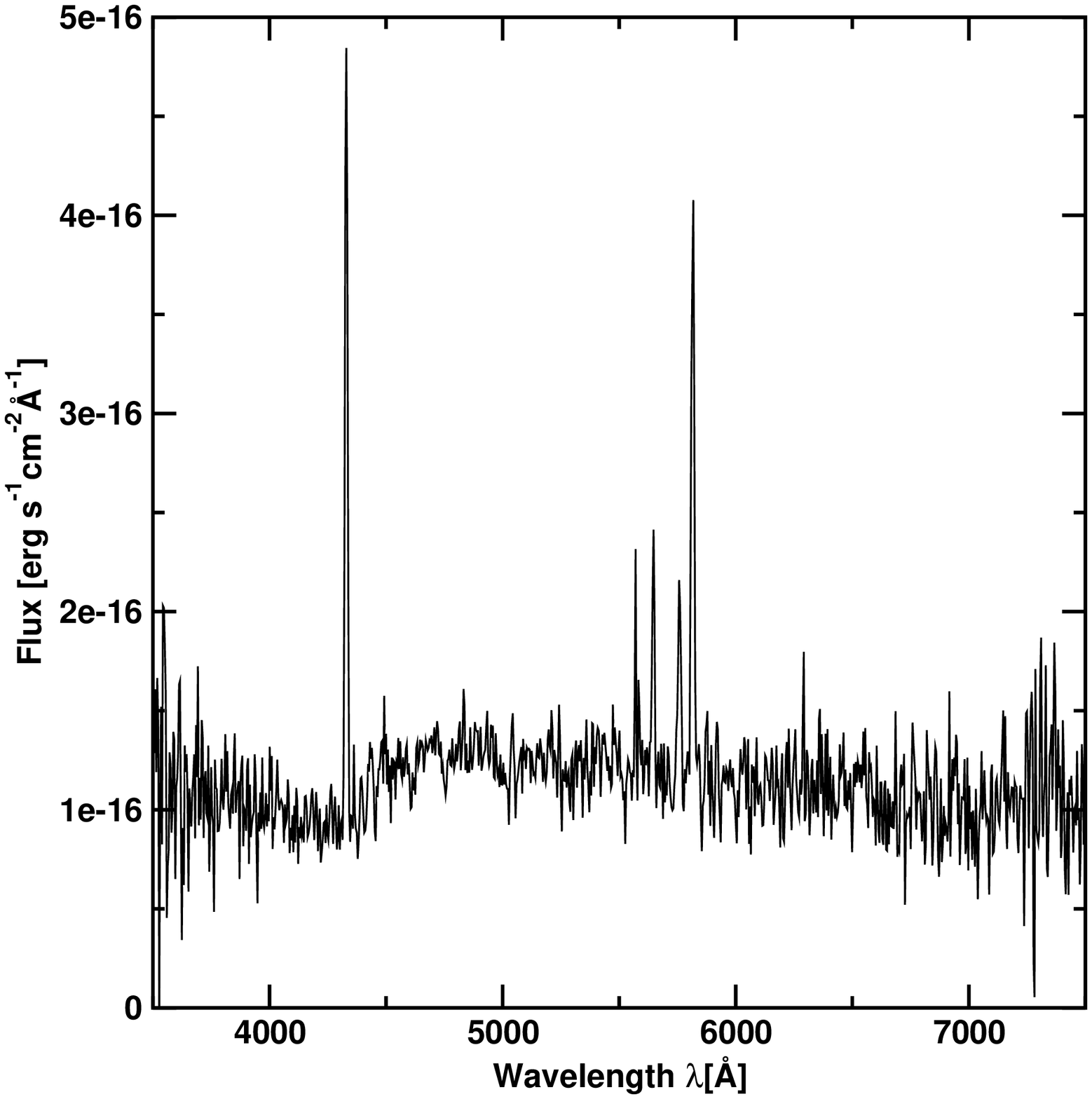}
\includegraphics[width=8cm]{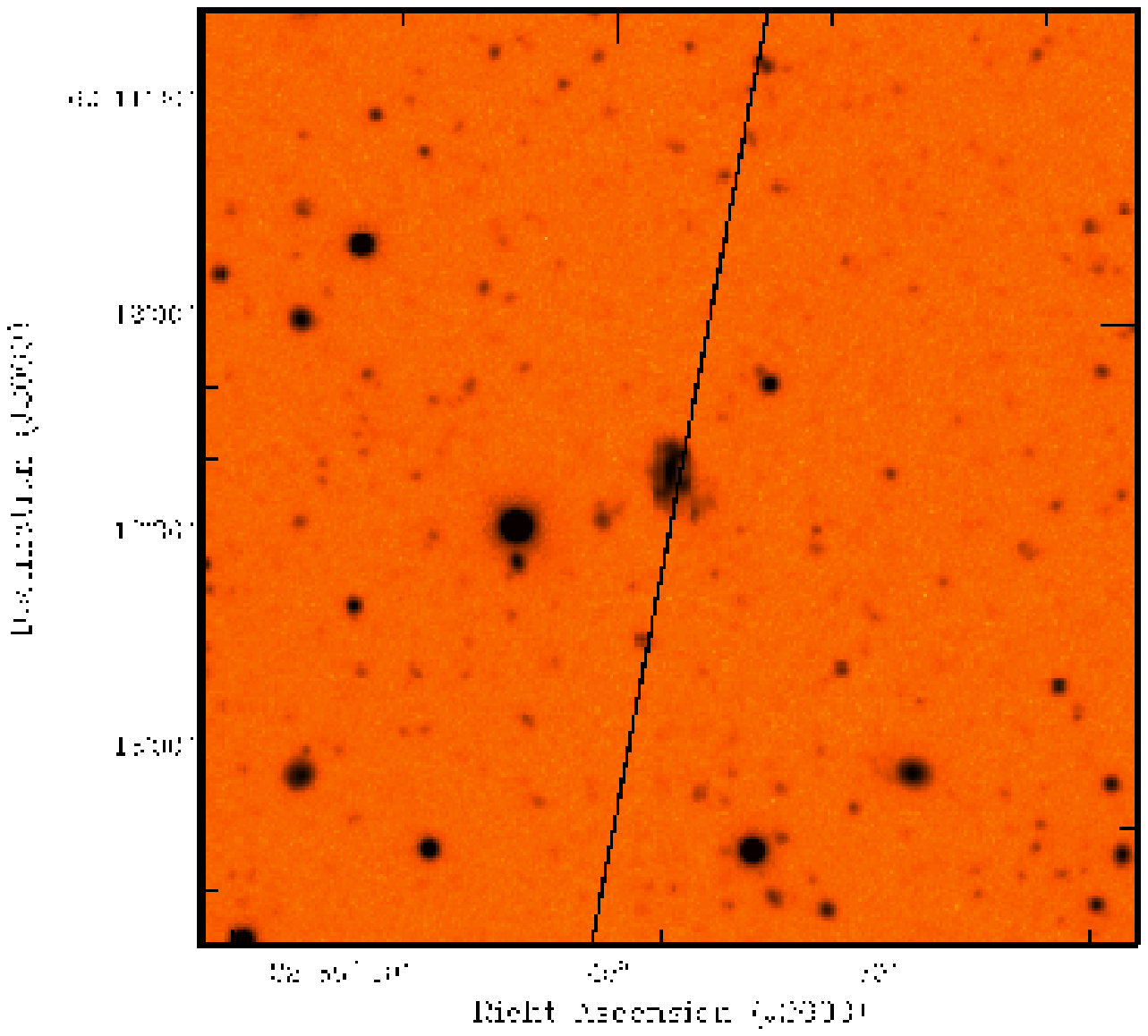}
\includegraphics[width=7.5cm]{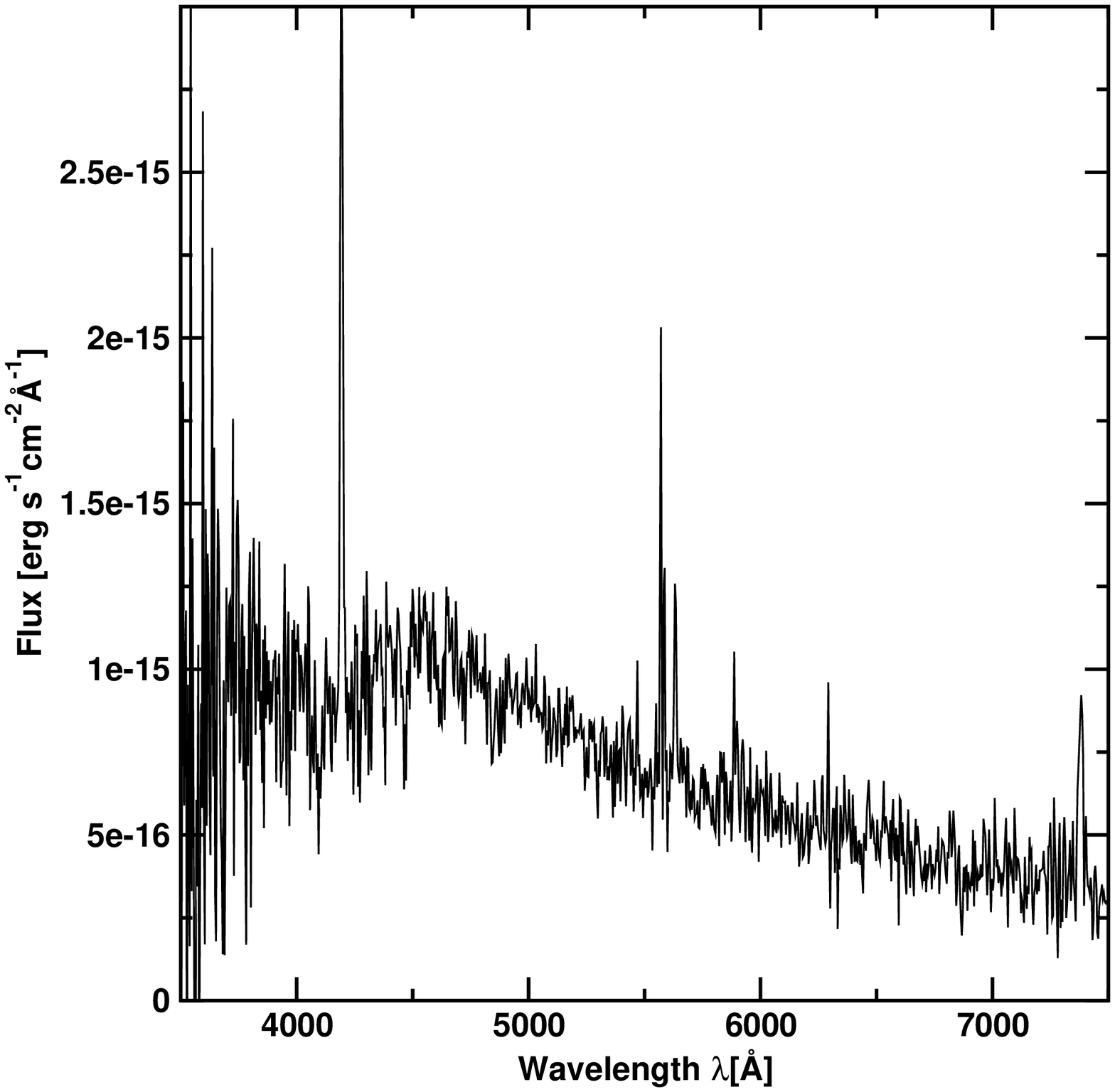}
\includegraphics[width=8cm]{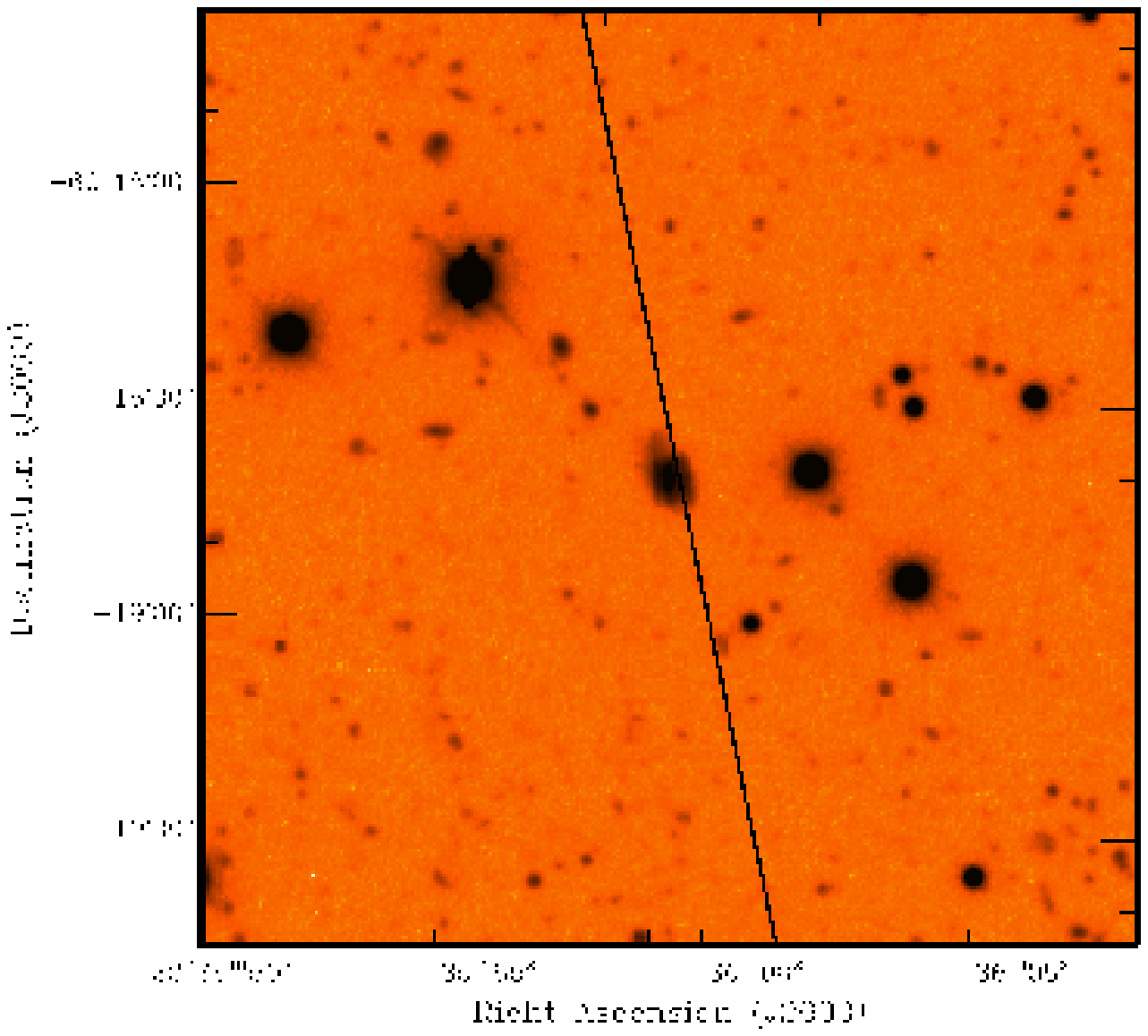}
\parbox{18cm}{{\bf Fig. 1 continued:} Form top to bottom:
  \object{LSB J22353-60311}, \object{LSB J22354-60122}, \object{LSB J22355-60183}}
\end{figure*}

\begin{acknowledgements}
This research was supported by the DFG Graduierten\-kolleg "The Magellanic
Systems, 
Galaxy Interaction and the Evolution of Dwarf Galaxies" (Universities
Bonn/Bochum). This work was funded by the DESY/BMBF grant 05 AE2PDA/8. 
This Paper is based on observations collected at the European Southern
Observatory, Chile Prog. Id. 66.A-0154(A). We thank the NOAO Deep Survey
team for making the pilot survey data immediately public, and the STIS team at
GSFC for the second data set. We also thank Jim Lauroesch for his help
  with the manuscript and the referee, who helped to improve the paper
  significantly.
\end{acknowledgements}

\bibliographystyle{aa}
\bibliography{6918bib}

\end{document}